\documentclass{article}

\usepackage{arxiv}

\usepackage[utf8]{inputenc} 
\usepackage[T1]{fontenc}    
\usepackage{hyperref}       
\usepackage{url}            
\usepackage{amsmath}
\usepackage{amssymb}
\usepackage{amsfonts}
\usepackage{booktabs}       
\usepackage{nicefrac}       
\usepackage{microtype}      
\usepackage{lipsum}
\usepackage{cite}

\usepackage{algorithmic}
\usepackage{graphicx}
\usepackage{subcaption}
\usepackage{textcomp}

\usepackage{soul}
\soulregister\cite7
\soulregister\citep7
\soulregister\citet7
\soulregister\ref7
\soulregister\pageref7
\soulregister\color7

\usepackage{color}

\title{A Novel Deep Neural Network Based Approach for Sparse Code Multiple Access}

\author{
  Jinzhi Lin\thanks{Corresponding author.} \\
  Shenzhen Institutes of Advanced Technology\\
  Chinese Academy of Science\\
  Shenzhen 518055, China\\
  \texttt{jz.lin@siat.ac.cn} \\
   \And
 Shengzhong Feng \\
  Shenzhen Institutes of Advanced Technology\\
  Chinese Academy of Science\\
  Shenzhen 518055, China\\
  \texttt{sz.feng@siat.ac.cn} \\
   \And
 Zhile Yang \\
  Shenzhen Institutes of Advanced Technology\\
  Chinese Academy of Science\\
  Shenzhen 518055, China\\
  \texttt{zl.yang@siat.ac.cn} \\
    \And
 Yun Zhang \\
  Shenzhen Institutes of Advanced Technology\\
  Chinese Academy of Science\\
  Shenzhen 518055, China\\
  \texttt{yun.zhang@siat.ac.cn} \\
   \And
 Yong Zhang \\
  Shenzhen Institutes of Advanced Technology\\
  Chinese Academy of Science\\
  Shenzhen 518055, China\\
  \texttt{zhangyong@siat.ac.cn} \\
}

\begin{document}
\maketitle

\begin{abstract}
Sparse code multiple access (SCMA) has been one of non-orthogonal multiple access (NOMA) schemes aiming to support high spectral efficiency and ubiquitous access requirements for 5G wireless communication networks. Conventional SCMA approaches are confronting remarkable challenges in designing low complexity high accuracy decoding algorithm and constructing optimum codebooks. Fortunately, the recent spotlighted deep learning technologies are of significant potentials in solving many communication engineering problems. Inspired by this, we explore approaches to improve SCMA performances with the help of deep learning methods. We propose and train a deep neural network (DNN) called DL-SCMA to learn to decode SCMA modulated signals corrupted by additive white Gaussian noise (AWGN). Putting encoding and decoding together, an autoencoder called AE-SCMA is established and trained to generate optimal SCMA codewords and reconstruct original bits. Furthermore, by manipulating the mapping vectors, an autoencoder is able to generalize SCMA, thus a dense code multiple access (DCMA) scheme is proposed. Simulations show that the DNN SCMA decoder significantly outperforms the conventional message passing algorithm (MPA) in terms of bit error rate (BER), symbol error rate (SER) and computational complexity, and AE-SCMA also demonstrates better performances via constructing better SCMA codebooks. The performance of deep learning aided DCMA is superior to the SCMA.
\end{abstract}


\section{Introduction}

To meet the emerging calls of massive connectivity and high spectral efficiency for modern mobile devices and Internet of Things, non-orthogonal multiple access (NOMA) is proposed in the 5th generation (5G) wireless communications. Unlike traditional orthogonal multiple access schemes, NOMA allows more than one user overlapping in the same communication resource block, causing interference while recovering the transmitted bits of all users by introducing a new demodulating algorithm. Typical NOMA schemes include power-domain NOMA\cite{huang2019optimal}, multi-user shared access (MUSA)\cite{yuan2016multi}, pattern division multiple access (PDMA)\cite{chen2017pattern}, bit division multiplexing (BDM)\cite{huang2014scalable}, interleave division multiple access (IDMA)\cite{ping2006interleave} and sparse code multiple access (SCMA)\cite{nikopour2013sparse}.

Inspired by low density signature (LDS) and multicarrier code division multiple access (CDMA)\cite{hoshyar2008novel}, SCMA merges QAM modulation and LDS spreader together: bits are grouped into sets, and each set is mapped into a complex sparse vector named codeword according to the predefined codebook. Different users are assigned to different codebooks, which are carefully designed such that non-orthogonal multiple access and overloading can be achieved. In a receiver's perspective, the received signal are "interleaved" by multiple users' transmitting signal. A maximum likelihood (ML) detector is supposed to obtain the optimal guess of transmitted bits from the received "interleaved" signal. Because of the sparsity of the codewords, the message passing algorithm (MPA) with lower computational complexity can be adopted to approximate the optimal solution of ML method.

There are two major challenges for implementation of an SCMA system: efficient decoding algorithm and optimal codebook design. Although MPA can be applied to calculate the marginal probabilities of all symbols based on an underlying factor graph, its iterative computation structure is too complicated and time-consuming for a practical detector. Finding a more efficient decoding algorithm for SCMA is an active topic in academia. As the codebook structure heavily dominates performance of the SCMA system, many researchers make effort to explore how to design better codebooks. Most of them consider methods of manipulating constellation structure to maximize the minimum Euclidean distance of constellation points inner and outer users. However, due to the non-orthogonality, more than one users collide over a carrier, thus introducing dependencies among different codewords can help to deduce to recover colliding codewords from the other carriers. As can be seen, it is hard to draw a criterion for constellation manipulating methods to guide to design good codebooks.

Thanks to the spotlighted breakthrough of AlphaGo\cite{silver2016mastering}, machine learning especially deep learning (DL) approaches have seen a dramatically increase of exciting applications in various fields such as computer vision and natural language processing. Generally speaking, DL is suitable for solving classification and recognition problems, particular in the areas that apparent physical features are not easy to characterize with rigid mathematical models. DL is good at extracting features for those systems by learning from a limited amount of labeled samples and categorizing the new unlabeled input data accordingly. Traditional technologies of modern communication contain rich expert knowledge of modeling, analyzing and designing methodology. However for SCMA, an obvious codebook design criterion and practical optimal decoder are intractable to obtain, and SCMA is essentially an encoding/decoding process with noise contamination. In this regard, DL has the large potential for SCMA in codebook generation and decoder construction.

In this paper, we are going to investigate how DL methods can be applied in SCMA systems and how well they can perform compared to the traditional and the state of the art approaches. To focus on these, throughout the paper, we assume that the discussed systems are all in the scenarios in which only AWGN channels without channel coding are considered.

\subsection{Related work}\label{SecRW}
Extensive works have been proposed in pursuing better and lower computational complexity SCMA decoding algorithms in recent years. Taking advantage of the linearly increasing complexity, Markov chain Monte Carlo (MCMC) method is applied for SCMA decoding in \cite{chen2018joint}, lower computational complexity is achieved when the codebook size is large. Authors in \cite{dai2017improved} design a look up table method to reduce the computational complexity of the MPA and propose several scheduling schemes achieving efficient message exchange and parallel processing to speed up the convergence of the MPA. By restricting function nodes in a reasonable search region and eliminating the exponential operations via applying appropriate combination of max operations, an improved log-MPA decoder called RRL detector\cite{tian2017low} demonstrates a near-optimum BER performance with significantly reduced complexity. An improved MPA which eliminates determined user codewords after certain number of iterations and continue the iterations for undetermined user’s codewords is proposed in \cite{Jia2018a}. \cite{wu2019low} has explored three low complexity detectors called variable MPA (VMPA), improved variable MPA (IVMPA) and incomplete iterative MPA (IIMPA) for reducing iteration times of the traditional MPA. A modified sphere decoding (MSD) detection scheme for SCMA is proposed in \cite{vameghestahbanati2017enabling}, which reduce complexity by exploiting the sparsity of the codebooks. This work achieves the performance of the optimal maximum likelihood (ML) detection in scenarios over AWGN channel without channel coding, thus is considered as the state of the art low-complexity SCMA detection compatible with our simulation setups. We choose it as a comparison candidate in the performance evaluation section.

In addition to the decoding algorithms, there are various studies focusing on SCMA codebook optimal designing. A systematic multi-stage lattice constellation based codebook design method is viewed as sub-optimal\cite{taherzadeh2014scma}. Various improvements, based on constellation rotation\cite{zhou2017scma}, spherical codebooks\cite{bao2016spherical}, star-QAM based multidimensional signaling\cite{yu2018design}, have been consecutively proposed. Taking the mapping matrices into account, \cite{peng2017joint} has presented a unified approach to generate constellations, which is a joint optimization problem formulated as a non-convex quadratically constrained quadratic programming that is tackled by using the semi-definite relaxation technique. By analyzing the SCMA signal model based on superposition modulation, \cite{dong2018efficient2} pointed out that the superimposed constellation points depend only on one amplitude variable, therefore proposed an SCMA codebook design method based on a one-dimensional searching algorithm, which can minimize the upper bound of pair-wise error probability (PEP) on the variable. Both of \cite{sharma2018scma} and \cite{dong2018efficient} have declared that mutual information can be utilized in designing SCMA codebooks, they respectively proposed different optimization methods based on that. Other SCMA codebook optimization methods, including genetic algorithm\cite{klimentyev2017scma}, constellation segmentation\cite{liu2018optimized}, dimensional permutation switching\cite{xiao2018capacity}, golden angle modulation\cite{mheich2019design} and maximum distance separable codes\cite{da2019multistage} have also been investigated.

All the above works are concerned with knowledge of conventional communication field and are independent of machine learning technologies. As DL exhibits its abilities in abundant fields\cite{LIU201711}, more and more researches are conducted for exploring its applicable possibility in communication systems. There are opportunities and challenges lie in these emerging studies\cite{qin2019deep}. New ways of thinking about communications as end-to-end reconstruction optimization tasks are introduced in \cite{o2017introduction}, which utilize autoencoders to jointly learn transmitter and receiver implementations as well as signal encodings without any prior knowledge. Similar thoughts are applied in OFDM\cite{felix2018ofdm}, massive MIMO systems\cite{huang2018deep}, millimeter-wave communications\cite{huang2019deepmmware}, optical fiber communications\cite{karanov2018end} and multi-colored visible light communications\cite{lee2018deep}. DL for channel coding is also attracting attentions\cite{nachmani2018deep, liang2018iterative}. In addition, many specific aspects of communication systems are being studied from machine learning perspective, including modulation recognition\cite{west2017deep}, PAPR reduction\cite{kim2018novel}, wireless interference identification\cite{schmidt2017wireless}, and so on. However, speaking to DL for SCMA, currently to the authors knowledge, very limited related works have been conducted or published besides the article \cite{kim2018deep}, which only takes account of autoencoders without DCMA extending.

\subsection{Paper contributions}
The major contributions of this paper can be summarized as follows:
\begin{enumerate}

\item A novel approach called DL-SCMA for SCMA decoding based on DL technique has been proposed. A deep neural network (DNN) model for learning how to decode SCMA is established, in where multiple-dimension vectors, which are converted from overlapped complex vectors derived from the received signal, are accepted as input, and binary vectors are output as the decoded bits.

\item An AE-SCMA scheme of designing autoencoders for SCMA en/decoding is established. The proposed AE-SCMA can consequently generate codebooks for the SCMA system and help to obtain the knowledge of the particular structure of optimal codebooks.

\item Viewed as a general version of SCMA, a novel dense code multiple access (DCMA) scheme is formulated, whose encoding and decoding processes are instructed and built by automatic learning of an autoencoder.

\end{enumerate}

The rest of this paper is organized as follows. Section \ref{SecModel} introduces the SCMA system model and architecture analysis of the neural network and autoencoder. Section \ref{SecSch} describes the proposed schemes. Experiments for performance evaluation and some discussions are conducted in Section \ref{SecEva} and Section \ref{SecDis} respectively. Section \ref{SecCon} concludes this paper and presents future work.

\section{SCMA system model and DL analysis} \label{SecModel}
\subsection{SCMA system model}

\begin{figure}
  \centering
    \includegraphics[width=0.8\textwidth]{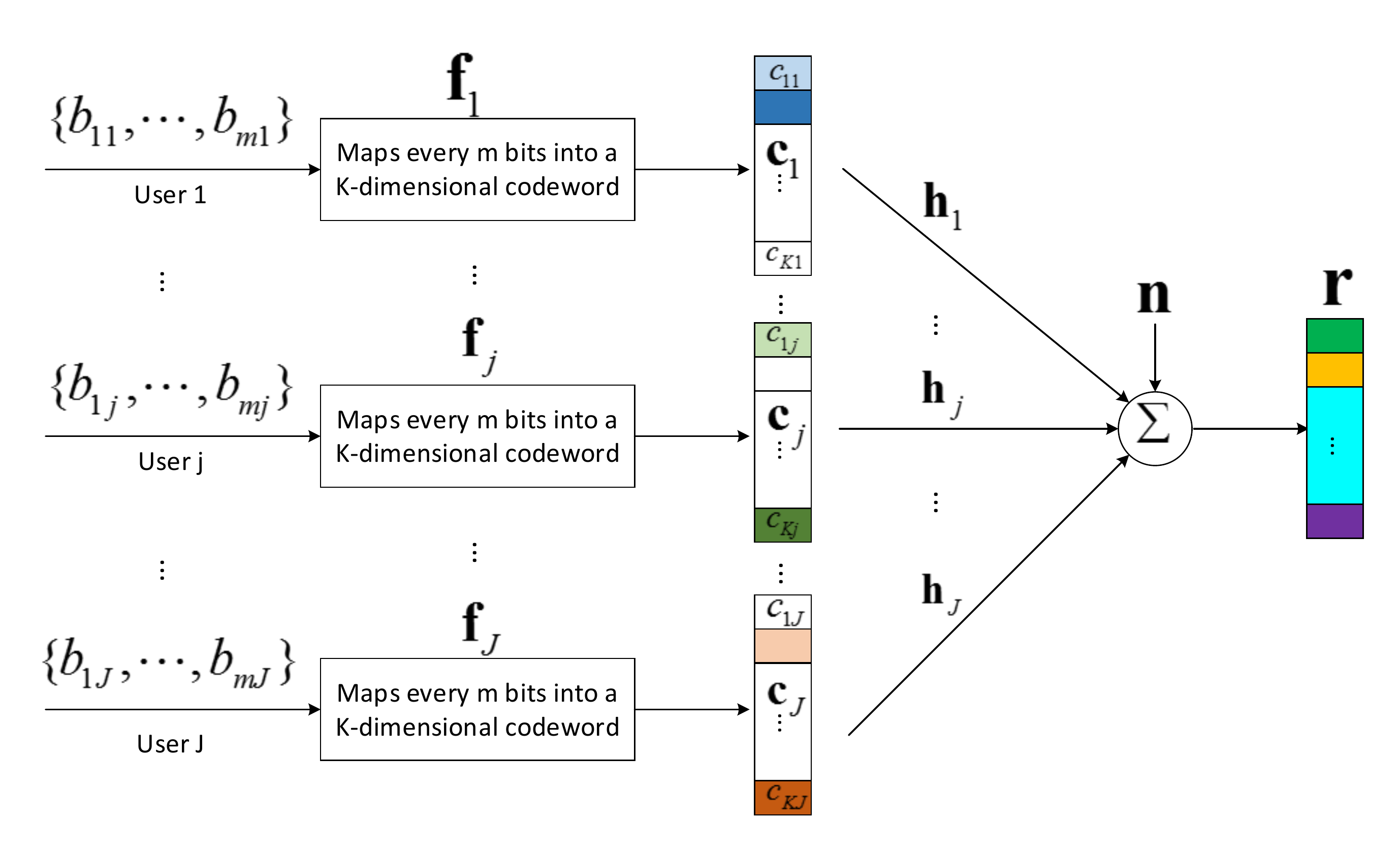}
  \caption{SCMA system model.}
  \label{FigSCMA}
\end{figure}

\begin{figure}
  \centering
    \includegraphics[width=0.8\textwidth]{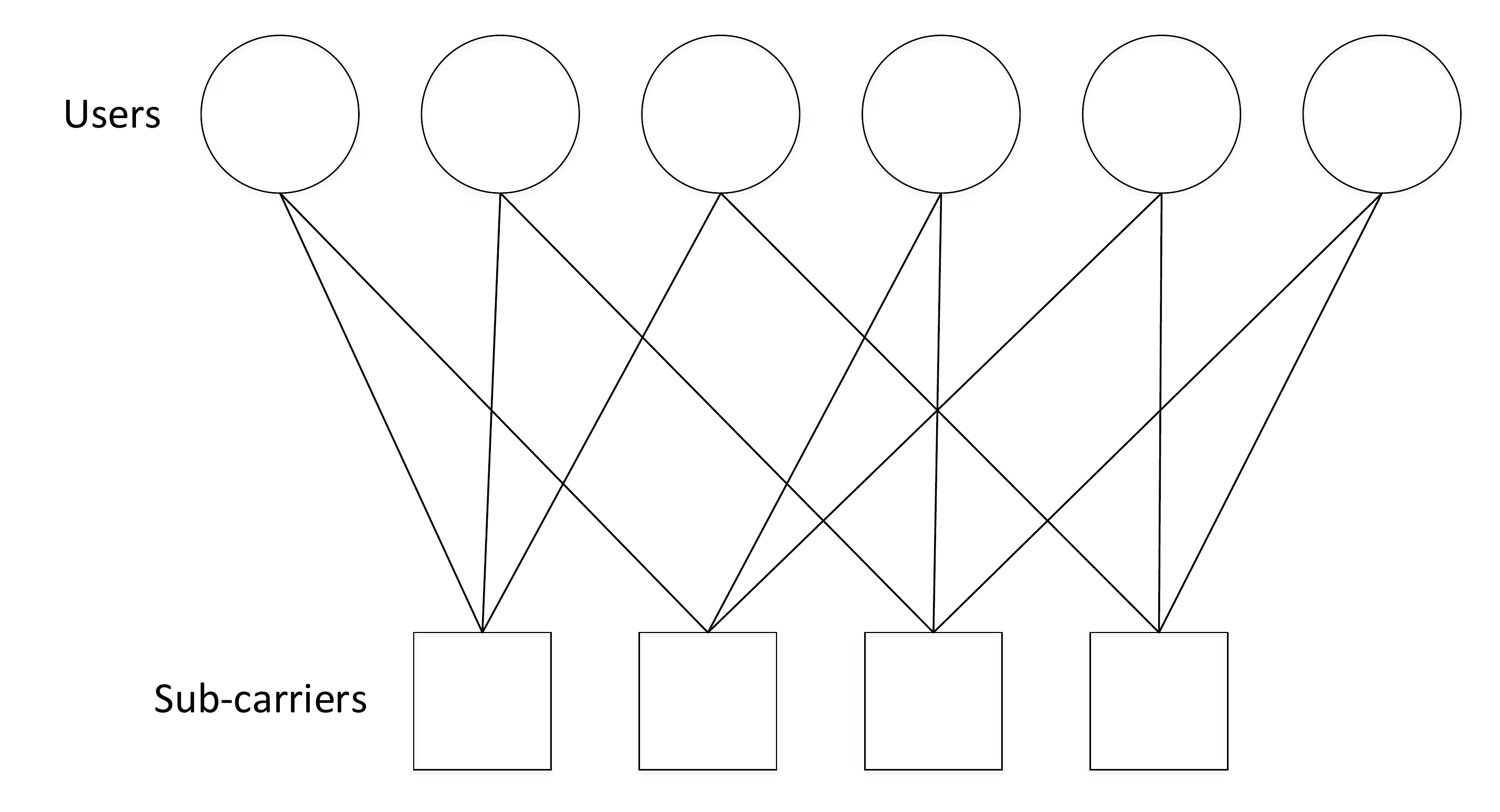}
  \caption{Factor graph representation of an SCMA encoder codeword mapping with 6 users and 4 sub-carriers.}
  \label{FigFactorGraph}
\end{figure}

As depicted in Fig. \ref{FigSCMA}, consider $J$ users transmitting data bits over the same $K$ sub-carriers of OFDM, here $K < J$ such that overloading is provided. According to SCMA encoder, each user maps every $m = {\log _2}(M)$ bits into a $K$-dimensional complex codeword $\boldsymbol{\bf{c}}$ with only $N$ non-zero elements standing for QAM modulation and LDS spreading combination, here $N < K$. The overlapping degree is $d_f = \frac{{JN}}{K}$, and overloading ratio is $\lambda  = \frac{J}{K}$. There are $M$ codewords forming a codebook for each user and each codebook is unique. The encoding procedure can be described by ${\boldsymbol{\bf{c}}} = {{\boldsymbol{\bf{f}}}}(\boldsymbol{\bf{b}})$, where $\boldsymbol{\bf{b}} \in \mathbb{B}^{{log_2}(M)}$ and  $\boldsymbol{\bf{c}} \in \mathcal{C} \subset \mathbb{C}^{K}$ with $\left| {\mathcal C} \right| = M$. Function $\boldsymbol{\bf{f}}$ is actually a mapping matrix which can be represented by a factor graph. Fig. \ref{FigFactorGraph} gives an example of the factor graph representation of 6 user data streams multiplexed over 4 sub-carriers.

Denote ${\boldsymbol{{\bf{b}}_j}} = \left( {{b_{1j}}, \cdots ,{b_{mj}}} \right)$, ${\boldsymbol{{\bf{c}}_j}} = {\left( {{c_{1j}}, \cdots ,{c_{Kj}}} \right)^T}$ and  ${\boldsymbol{{\bf{f}}_j}} = {\left( {{f_{1j}}, \cdots ,{f_{Kj}}} \right)^T}$ as the transmitting bits, mapped codeword and  the mapping functions of user $j$, respectively. After synchronous multiplexing, without considering channel fading, the received signal can be expressed as:
\begin{equation}
\begin{split}
{\boldsymbol{\bf{r}}} & = \sum\nolimits_{j = 1}^J {diag({{\boldsymbol{\bf{h}}}_j}){{\boldsymbol{\bf{c}}}_j}}  + {\boldsymbol{\bf{n}}}\\
 & = \sum\nolimits_{j = 1}^J {diag({{\boldsymbol{\bf{h}}}_j}){{\boldsymbol{\bf{f}}}_j}({{\boldsymbol{\bf{b}}}_j})}  + {\boldsymbol{\bf{n}}}
\end{split}
\end{equation}
where ${\boldsymbol{{\bf{h}}_j}} = {\left( {{h_{1j}}, \cdots ,{h_{Kj}}} \right)^T}$ is the channel gain vector for $K$ sub-carriers of user $j$, here all its elements are set to constants as no channel fading is considered, and ${\boldsymbol{\bf{n}}} = {\left( {{n_{1j}}, \cdots ,{n_{Kj}}} \right)^T},{n_{ij}} \sim {\mathcal C}{\mathcal N}\left( {0,{\sigma ^2}} \right)$ is the additive white Gaussian noise (AWGN) vector. The key variables that affects the received signal are the codewords ${{\boldsymbol{\bf{c}}}_j}$ which are determined by the mapping functions ${{\boldsymbol{\bf{f}}}_j},j = 1, \cdots J$. 

For an SCMA decoder, its mission is to recover the original bits transmitted by all the users as far as possible given the received signal $\boldsymbol{\bf{r}}$, channel conditions $\left\{ {{{\boldsymbol{\bf{h}}}_j}} \right\}_{j = 1}^J$ and all the user codewords $\left\{ {{{\mathcal C}_j}} \right\}_{j = 1}^J$. The joint optimum maximum \textit{a posteriori} (MAP) detection can guess a $\hat {\boldsymbol{\bf{C}}}$ that maximizes the joint \textit{a posteriori} pmf (probability mass function) of the multiplexed codewords ${\boldsymbol{\bf{C}}} \in {{\mathcal C}^J}$, which can be expressed as:
\begin{equation}
\hat {\boldsymbol{\bf{C}}} = \arg \mathop {\max }\limits_{{\boldsymbol{\bf{C}}} \in {{\mathcal C}^J}} p({\boldsymbol{\bf{C}}}|{\boldsymbol{\bf{r}}})
\end{equation}
where ${{\mathcal C}^J}: = {{\mathcal C}_1} \times  \cdots  \times {{\mathcal C}_J}$. The complexity of MAP detector increases exponentially with $J$ and polynomially with $M$, thus the MPA detector, which interactively approximates the solution of the marginalize product of functions (MPF) problem over the underlying factor graph, is applied as a near-optimal solution.

\subsection{Deep neural network}

\begin{figure}
  \centering
    \includegraphics[width=0.8\textwidth]{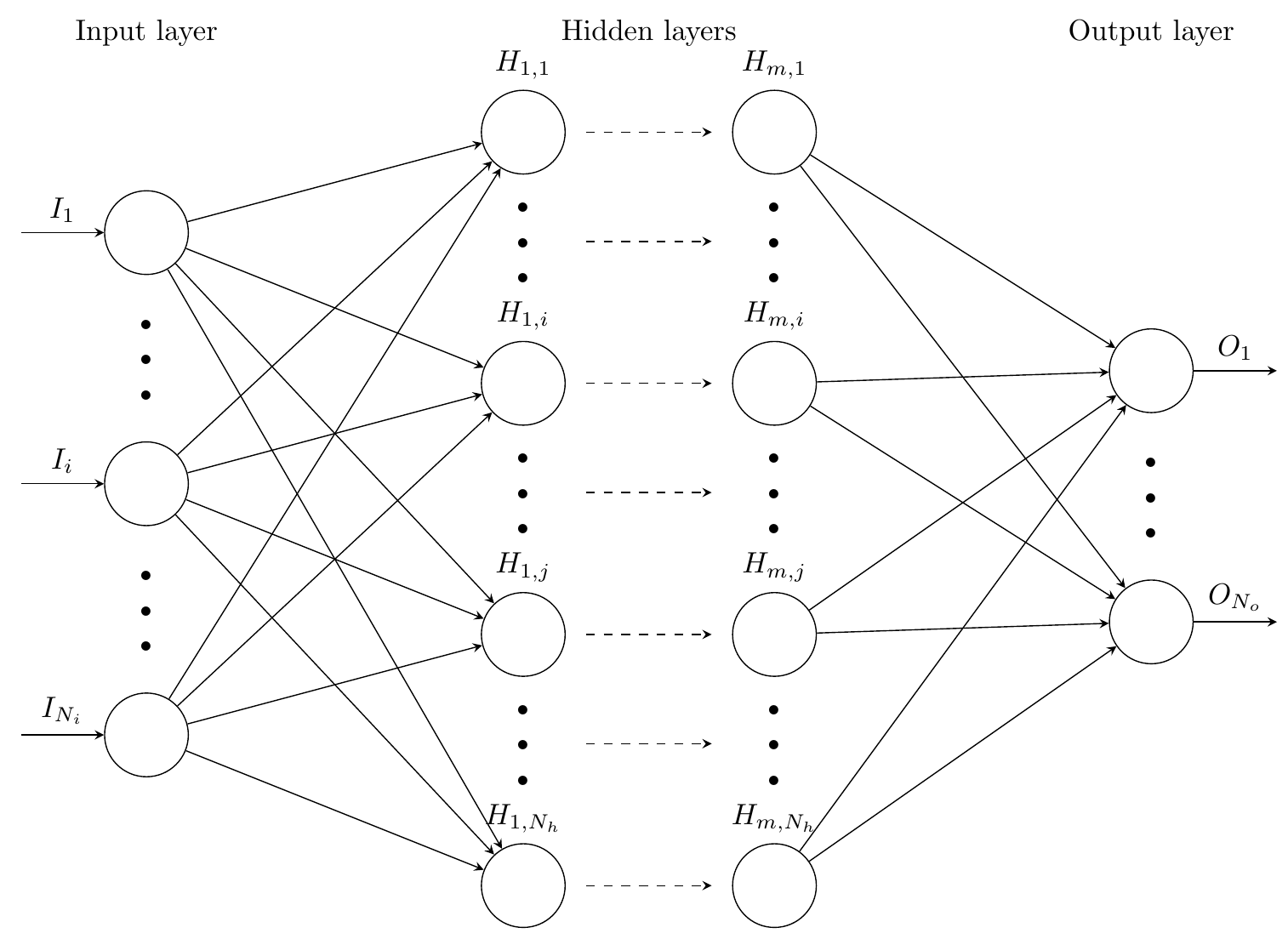}
  \caption{Structure of a DNN.}
  \label{FigDNN}
\end{figure}

A DNN is composed of multiple layers which are made of nodes called "neurons". A node is the place where computation happens, which is described by:
\begin{equation}
y = \varphi ({\boldsymbol{{\bf{w}}}^T}\boldsymbol{{\bf{x}}} + b)
\end{equation}
where $\boldsymbol{{\bf{w}}} = ({w_1}, \cdots ,{w_n})^T \in \mathbb{R}^n, \boldsymbol{{\bf{x}}} = ({x_1}, \cdots ,{x_n}) ^T\in \mathbb{R}^n, b \in \mathbb{R}$. Each node in a layer accepts all the output data of the previous layer as input $\boldsymbol{\bf{x}}$. Each input data $x_i$ is multiplied by a weight $w_i$, and all the multiplied data plus a bias $b$ are added up and the sum is passed through an activation function $\varphi$ to generate the output $y$. Apart from the input and output layers, a DNN usually has more than one hidden layers. Each layer $l$ with $N_{l,o}$ nodes connecting a preceding layer with $N_{l,i}$ nodes can be described by: 
\begin{equation} \label{EqLayer}
{{\boldsymbol{\bf{y}}}_l} = {\varphi _l}\left( {{{\boldsymbol{\bf{W}}}_l}^T{{\boldsymbol{\bf{x}}}_l} + {{\boldsymbol{\bf{b}}}_l}} \right)
\end{equation}
where ${{\boldsymbol{\bf{W}}}_l} \in {{\mathbb{R}}^{{N_{l,i}} \times {N_{l,o}}}}$ is the weight matrix, ${\boldsymbol{\bf{b}}}_l  \in {\mathbb{R}^{N_{l,o}}}$ is the bias vector, and ${\boldsymbol{\bf{x}}}_l  \in {\mathbb{R}^{N_{l,i}}}$, ${\boldsymbol{\bf{y}}_l}  \in {\mathbb{R}^{N_{l,o}}}$ are the input and output vectors, respectively. The structure of a DNN is shown in Fig. \ref{FigDNN}. 

To train a DNN, back propagation and gradient descent are commonly used approaches. Firstly, a loss function $L(\cdot)$ which calculates the difference between the network output and its expected output is needed. Normally, the mean squared error is used for the loss function. In the case that output data are vectors of binary or probabilities whose values are in the range of $[0,1]$, then the cross-entropy can be adopted as a better loss function. Then, based on the gradient descent method, the weights and biases of the network are updated according to their derivatives of the loss function, which is expressed as follows:
\begin{equation}
\begin{array}{c}
{w_{ij}}^\prime  = {w_{ij}} - \alpha \frac{{\partial L}}{{\partial {w_{ij}}}} = {w_{ij}} - \alpha \frac{{\partial L}}{{\partial {y_j}}}\frac{{\partial {y_j}}}{{\partial ne{t_j}}}{y_i}\\
{b_{ij}}^\prime  = {b_{ij}} - \alpha \frac{{\partial L}}{{\partial {b_{ij}}}} = {b_{ij}} - \alpha \frac{{\partial L}}{{\partial {y_j}}}\frac{{\partial {y_j}}}{{\partial ne{t_j}}}
\end{array}
\end{equation}
where $\alpha$ is the learning rates and $\frac{{\partial {y_j}}}{{\partial ne{t_j}}}$ is the partial derivative of the activation function of layer $j$ with respective to its input. To gain better training performance and decrease computational complexity, many techniques for improving the gradient descent method are proposed, such as: SGD, Momentum, Adagrad, RMSprop, Adam, \textit{et al.}.

\subsection{Autoencoder}
\begin{figure}
  \centering
    \includegraphics[width=0.8\textwidth]{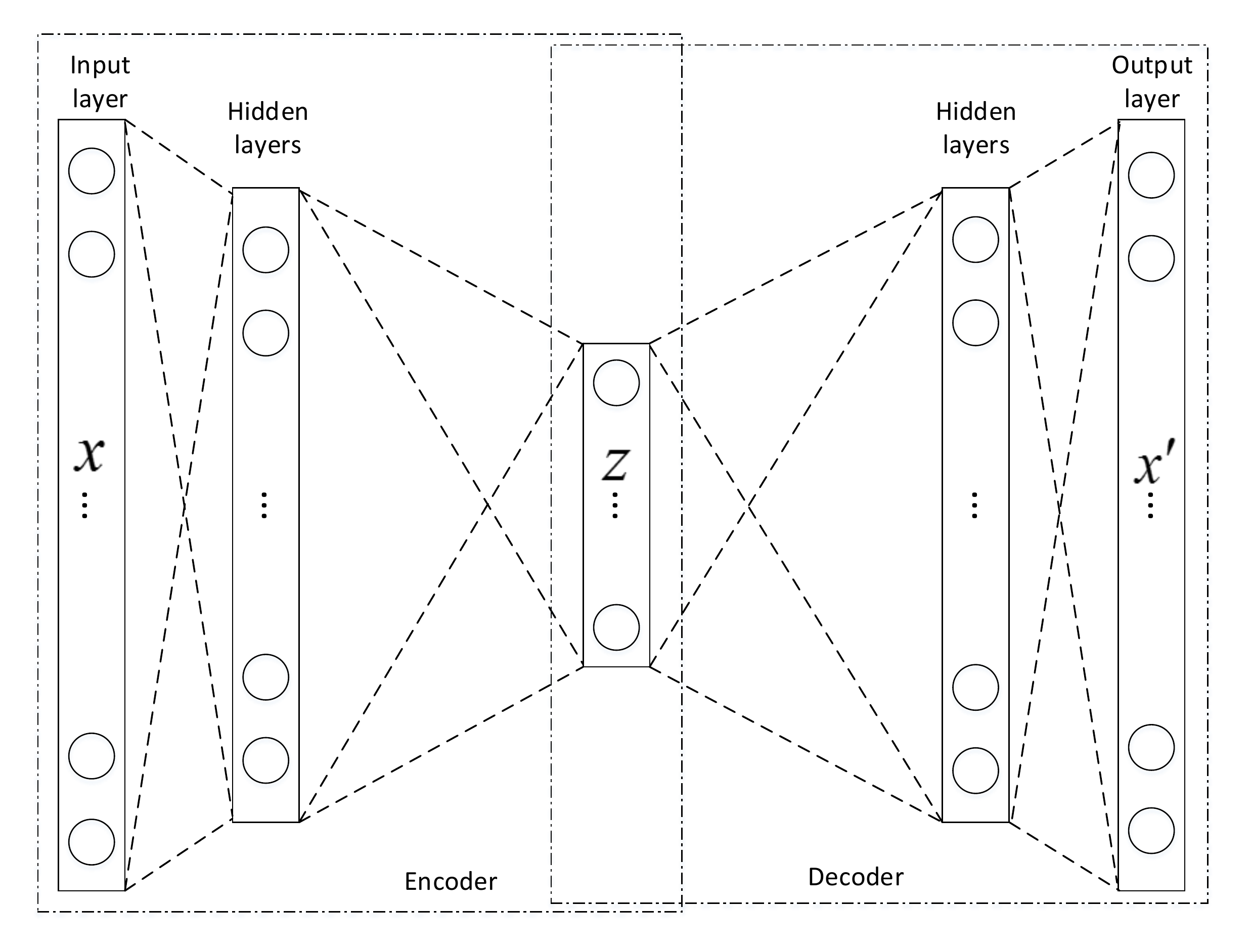}
  \caption{Structure of an autoencoder.}
  \label{FigAEArch}
\end{figure}

Autoencoders can be viewed as special neural networks that their output values are equal to the inputs. It is composed of two parts: the encoder that learns to compress data from the input layer into a code and the decoder that learns to uncompress the code into values which closely matches the original input data. The structure of an autoencoder is depicted in Fig. \ref{FigAEArch}.

As one of the autoencoder's variants, denoising autoencoders (DAE) deal with inputs corrupted by some form of noise and are trained to recover the original clean versions\cite{liu201880}. Suppose $C( \cdot )$ is a corruption process, the input of a DAE can be expressed as $\tilde {\boldsymbol{\bf{x}}} = C(\boldsymbol{{\bf{x}}})$. Denote $e( \cdot ;{{\boldsymbol{\bf{W}}}_e},{\boldsymbol{{\bf{b}}}_e})$ and $d( \cdot ;{{\boldsymbol{\bf{W}}}_d},{{\boldsymbol{\bf{b}}}_d})$ as the encoder and decoder, respectively, where ${{\boldsymbol{\bf{W}}}_e},{{\boldsymbol{\bf{b}}}_e},{{\boldsymbol{\bf{W}}}_d},{{\boldsymbol{\bf{b}}}_d}$ are the weights and biases to be determined, and the target of training the DAE is to minimize the reconstruction loss:
\begin{equation} \label{EqDAE}
{\bf{W}}_e^*,{\bf{b}}_e^*,{\bf{W}}_d^*,{\bf{b}}_d^* = \mathop {\arg \min }\limits_{{{\bf{W}}_{e,d}},{{\bf{b}}_{e,d}}} L\left( {{\bf{x}},d\left( {e\left( {{\bf{\tilde x}};{{\bf{W}}_e},{{\bf{b}}_e}} \right);{{\bf{W}}_d},{{\bf{b}}_d}} \right)} \right)
\end{equation}
Denoising training forces $e( \cdot ;{{{\boldsymbol{\bf{W}}}_e},{{\boldsymbol{\bf{b}}}_e}})$ and $d( \cdot ;{{{\boldsymbol{\bf{W}}}_d},{{\boldsymbol{\bf{b}}}_d}})$ to implicitly learn the structure of the corrupted input samples, so as to gain a good representation of the general input model.

\section{Proposed schemes} \label{SecSch}
In this section, implementation of DL-SCMA based on DNN is firstly presented. Then AE-SCMA schemes of applying autoencoders to learn to do SCMA coding and decoding are proposed. In addition, techniques of variable initialization and batch normalization for the proposed schemes have been discussed.

\subsection{DNN based SCMA decoder}
As given in the previous section, in SCMA, the received signal $\boldsymbol{\bf{r}}$ is the combination of multiple users' modulated signals plus channel noise. In this paper, only AWGN is considered, and assume all channel conditions  $\left\{ {{{\boldsymbol{\bf{h}}}_j}} \right\}_{j = 1}^J$ are known and determined. Viewed from ends of transmitters and receivers, different transmitting bit combinations should generate different received signals, that's to say, the transmitting bits and the received signal are correlated with certain relations. Generally speaking, because of the participation of noises, these relations are not clear and therefore difficult to formulate an explicit mathematics equation to describe without enough knowledge of users' codebooks and channel conditions. For a receiver, its task is to find these relations. This is a typical classification problem: classify different received signals into the corresponding transmitting bit combinations. It is well known that DNNs are good at learning features of implicit relations and solve classification problems. Therefore, applying DNNs to implement SCMA decoders is feasible.

For an SCMA system of $J$ users transmitting bits over $K$ sub-carriers of OFDM, a DNN for SCMA decoding accepts the received signal which can be represented by $K$ complex values corrupted by AWGN noise, and output $mJ$ binaries that stand for the decoded bits transmitted by the $J$ users, where $m = {\log _2}(M)$, given previously. Converting the complex values into real and imaginary parts, an SCMA decoding DNN has $2K$ nodes for the input layer and $mJ$ nodes for the output layer. Hidden layers are dense fully connected, whose activation functions could be rectified linear unit (ReLU), \textit{Tanh} or \textit{Sigmoid}. Through running many experiments in practice, the \textit{Tanh} activation function is proved to be the best in gaining better performance for SCMA decoding. To balance the decoding performance and computational complexity, the number of hidden layers and number of nodes for each hidden layer should be determined carefully.

Generally, to solve a classification problem, a DNN's output layer usually produces probabilities of multi-classes ${p_i},i = 1, \cdots ,n$, and cross entropy is usually chosen as the natural loss function $L$ defined as:
\begin{equation}
L =  - \sum\nolimits_{i = 1}^n {{t_i}\log ({p_i})} 
\end{equation}
where $t_i,i = 1, \cdots ,n$ are the target probabilities, which typically take values of one or zero, and they are the supervised information provided to train the DNN classifier. 

There are two cases in computing output probabilities: for mutually exclusive classification, where classified labels must be one-hot encoded or form a soft class probability distribution, \textit{Softmax} function should be used because it can combine all the input elements together to transform them into a class probability distribution. While, for independent non-mutually exclusive classification, multiple classes can coexist in one classification, and \textit{Sigmoid} function is a better choice, for that it can compute every input element's probability individually. As for SCMA decoding, outputs are binary bits that are independent non-mutually exclusive. So, for the SCMA decoding DNN, \textit{Sigmoid} function $\sigma (x) \equiv {(1 + {e^{ - x}})^{ - 1}}$ is used as the activation function in the output layer, and cross entropy is adopted as the loss function.
 
The aim of training this SCMA decoding DNN is to approximate the optimum solution of weights and biases, ${\boldsymbol{\bf{W}}}_d,{\boldsymbol{\bf{b}}}_d$, to minimize the cross entropy between the original transmitting bits from all users and the DNN outputs, which can be expressed as following:
\begin{equation} \label{EqDNN}
\mathop {\min }\limits_{{{\boldsymbol{\bf{W}}}_d},{{\boldsymbol{\bf{b}}}_d}} \left( { - \sum\limits_{i = 1}^{mJ} {{b_i}\log \left( {{\pi _i}\left[ {d\left( {{{\left( {{y_1}, \cdots ,{y_{2K}}} \right)}^T};{{\boldsymbol{\bf{W}}}_d},{{\boldsymbol{\bf{b}}}_d}} \right)} \right]} \right)} } \right)
\end{equation}
where $d( \cdot ;{\boldsymbol{\bf{W}}}_d,{\boldsymbol{\bf{b}}}_d)$ represents the whole neural network decoding process which outputs the decoded bits vector,  $\hat {\boldsymbol{\bf{b}}} = {\left( {{{\hat b}_1}, \cdots ,{{\hat b}_{mJ}}} \right)^T}$. Note that, ${{\hat b}_i},i = 1, \cdots ,mJ$ are the outputs of \textit{Sigmoid} function, whose values are in the range of $[0,1]$. ${\pi _i}[ \cdot ]$ denotes the $i$-th element of a input vector.

To train the network, sufficient samples containing received signal ${{\boldsymbol{\bf{y}}}}$ and the corresponding original transmitting bits ${{\boldsymbol{\bf{b}}}}$, where ${\boldsymbol{\bf{y}}} = {{{\left( {{y_1}, \cdots ,{y_{2K}}} \right)}^T}}$ and  ${\boldsymbol{\bf{b}}} = {\left( {{b_1}, \cdots ,{b_{mJ}}} \right)^T}$, are required. To get sufficient knowledge of users' modulating codewords, enough various combinations of ${\boldsymbol{\bf{b}}}$ should be provided in the training sample set. For this reason, in this paper, ${{\boldsymbol{\bf{b}}}}$ are randomly generated in uniform distribution for the training sample set. 

Signal-to-noise ratio (SNR) or ${E_b}/{N_0}$ plays an important role in communication systems and usually directly affects the BER. It can be inferred that training samples with different SNRs may lead to different learning results for the SCMA decoding DNN. Too small SNR means that the original modulated signal can be severely corrupted by noise, causing signal structure vague. Thus it is of significant difficulties for the DNN training to extract samples' inner features and learn to decode correctly. On the other hand, training samples generated by too large SNR could lead to overfitting, thereby causing bad performance when dealing with small SNR testing data sets. Overall, appropriate SNR to generate training samples is crucial for the decoding performance and should be investigated comprehensively.

\subsection{Learn to en/decode}
It is notable that, with only the aid of sufficient training samples, DL-SCMA can learn to decode SCMA without any knowledge of users' codewords ${{\boldsymbol{\bf{c}}}_j}$ and channel conditions ${{\boldsymbol{\bf{h}}}_j},j = 1, \cdots J$. That's to say, codebooks and channel information can be learned and contained implicitly in the weights and biases of the network. Actually, not only can the decoding process be constructed by a DNN, but also can the encoding process be done so. Senders of an SCMA system map every $m$ bits to a $K$-dimensional complex value vector (codeword). A DNN with $m$ nodes of input layer accepting binaries and $2K$ nodes of output layer standing for the $K$-dimensional complex codeword can represent the SCMA mapping (encoding) process for one single user. The multiple users synchronously transmitting process are represented by stacking up $J$ such DNNs and summing up their output layers. Viewed these stacked DNNs as a whole, it could be called SCMA DNN encoder and connected with the aforementioned decoding DNN. Hence, an autoencoder is obtained. 

\begin{figure*}
  \centering
    \includegraphics[width=1.0\textwidth]{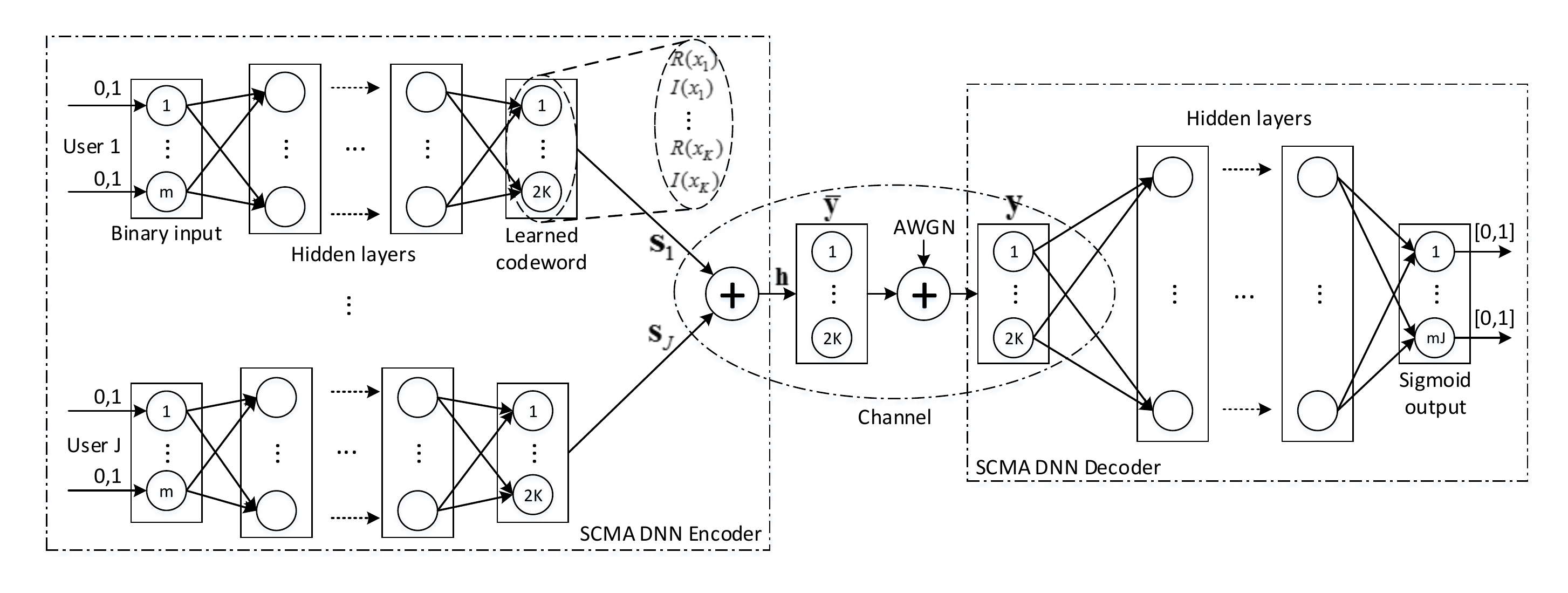}
  \caption{Structure of the SCMA autoencoder.}
  \label{FigSCMAAE}
\end{figure*}

Fig. \ref{FigSCMAAE} gives the whole detailed structure of the SCMA autoencoder. To keep the learned codewords sparse, a binary vector ${{\boldsymbol{\bf{s}}}_j} = {\left( {{s_1}, \cdots ,{s_{2K}}} \right)^T}$ which is determined by the SCMA mapping matrix is multiplied to the encoder's output layer for each user $j$. Elements of ${\boldsymbol{\bf{s}}}_j$ taking values of 1 or 0 depend on which $N$ resources the user $j$ is mapped to. For example, for case $J=6$, $K=4$ and $N=2$, if user $j$ is mapped to resources 2 and 4, then $s_j$ takes the value of $[0,0,1,1,0,0,1,1]^T$. The output signal in resources 1 and 3 are canceled out after being multiplied by vector $s_j$. By doing so, although every user output $2K$ nodes, only $2N$ of them have none-zero values. By summing up all the users' outputs, the combined output signal is the same as the original SCMA system.
In addition, a vector ${\boldsymbol{\bf{h}}}$ stands for the channel conditions is multiplied to the sum of the $J$ users' encoder output layers, and an AWGN vector is added to the sum layer ${{\boldsymbol{\bf{\bar y}}}}$ to produce the decoder input layer ${\boldsymbol{\bf{y}}}$. Here, all the $2K$ elements of vector ${\boldsymbol{\bf{h}}}$ are set to constant values representing the complex channel gains for the $K$ sub-carries travelling in the assumed AWGN channels. Stacking multiple DNNs together and connecting with another DNN is easy to implement in Tensorflow. Furthermore, experiments have shown that, for such an autoencoder with noise added, back-propagation based training works without problem in Tensorflow.

The same as a normal autoencoder, AE-SCMA is composed of two parts: the encoder and decoder. A small difference lies in that communication channel and noise are introduced connecting the encoder and the decoder. As can be seen, this is actually a DAE that the corruption process is dominated by AWGN in the channel. Outputs of the encoder part stand for the learned codewords, including real and imaginary parts. To constrain signal's transmitting energy, the \textit{Tanh} activation function is adopted in the output layer of the encoder part, causing values of the real and imaginary parts of the learned codewords in the range of [-1,1], thus the maximum power is 2. The AWGN is randomly generated by a normal distribution with mean equals zero and variance ${\sigma ^2} = E\left[ {{{\left\| {{\boldsymbol{\bf{\bar y}}}} \right\|}^2}} \right]/SNR$, where $E[ \cdot ]$ is the expectation value and $SNR$ is calculated by $SNR = ({E_b}/{N_0}) \cdot (mJ/K)$. 

Similar to DL-SCMA, due to that only binaries are accepted in the input layer, cross-entropy should be chosen as the loss function for training and \textit{Sigmoid} function should be used as the activation function of the decoder's output layer. Still, experiments prove that \textit{Tanh} function performs better than other activation functions for the hidden layers in AE-SCMA. The number of layers and the number of nodes for each layer are carefully designed according to the balance of the desired performance and computational complexity.

Shown in Fig. \ref{FigSCMAAE}, the binary vector ${\boldsymbol{\bf{s}}}_j$ whose element values can only be 0 or 1 indicates which sub-carriers are occupied by user $j$. Vectors ${\boldsymbol{\bf{s}}}_j,j=1, \cdots ,J$ are not trainable, and they should be configured as constants according to the SCMA mapping matrix before training. For a regular SCMA system, there are only $N$ non-zero elements within a user's codeword, thus, correspondingly, there are $2N$ elements take value 1 in ${\boldsymbol{\bf{s}}}_j$. As a result, the outputs of the encoder part of AE-SCMA can be viewed as codewords which are learned by training. 

As we all know, the reason why SCMA is called \textit{spare} is that it is a relatively small percentage of elements taking value 1 in the codewords, that is to say $N \ll K$. However, in AE-SCMA, by setting all or most of the values of elements in ${\boldsymbol{\bf{s}}}_j$ to 1, we get a new form of code based NOMA scheme called \textit{dense code multiple access} (DCMA). Compare to SCMA, DCMA has a higher overlapping degree, $d_f$. For AE-SCMA, it treats SCMA and DCMA with no difference, as its network structure and training procedure do not change.

According to (\ref{EqDAE}) and (\ref{EqDNN}), the denoising training target of AE-SCMA is given as following:
\begin{equation} 
\mathop {\min }\limits_{{{\boldsymbol{\bf{W}}}_{e,d}},{{\boldsymbol{\bf{b}}}_{e,d}}} \left( { - \sum\limits_{i = i}^{mJ} {{b_i}\log \left( {{\pi _i}\left[ {d\left( {{\boldsymbol{\bf{h}}} \cdot e({\boldsymbol{\bf{b}}};{{\boldsymbol{\bf{W}}}_e},{{\boldsymbol{\bf{b}}}_e}) + {\boldsymbol{\bf{n}}};{{\boldsymbol{\bf{W}}}_d},{{\boldsymbol{\bf{b}}}_d}} \right)} \right]} \right)} } \right)
\end{equation} 
where $e({\boldsymbol{\bf{b}}};{{\boldsymbol{\bf{W}}}_e},{{\boldsymbol{\bf{b}}}_e}) = \sum\nolimits_{j = 1}^J {{{\boldsymbol{\bf{s}}}_j} \cdot {e_j}\left( {{{\boldsymbol{\bf{b}}}_j};{{\boldsymbol{\bf{W}}}_e}^j,{{\boldsymbol{\bf{b}}}_e}^j} \right)}$ and $\boldsymbol{\bf{n}}$ is the vector of AWGN related to $SNR$. In order to make AE-SCMA to learn to construct complete codebooks for all users, training set should contain all the $M^J$ combinations of ${\boldsymbol{\bf{b}}}$. Similar to the SCMA decoding DNN, an appropriate value of $SNR$ or ${E_b}/{N_0}$ should be considered carefully in the training process as it is crucial for AE-SCMA to extract essential features from the training data.

\subsection{Variable initialization and batch normalization}
Due to back propagation and gradient descent based training, different initial values of weights and biases may lead a DNN to different finial solutions and convergence speeds\cite{CAO2018278}. Variable initialization is important for both of the SCMA decoder DNN and autoencoder proposed in this paper.

Activation function \textit{Tanh} is used for the hidden layers of the SCMA decoding DNN and autoencoder. According to (\ref{EqLayer}), for each hidden layer $l$, its output  ${{\boldsymbol{\bf{y}}}_l} = {({y_{l,1}}, \cdots {y_{l,{N_{l,o}}}})^T}$ is dominated by:
\begin{equation} 
{{\boldsymbol{\bf{y}}}_l} = Tanh({{\boldsymbol{\bf{W}}}_l}^T{{\boldsymbol{\bf{y}}}_{l - 1}} + {{\boldsymbol{\bf{b}}}_l})
\end{equation} 
where ${{\boldsymbol{\bf{y}}}_{l - 1}} = {({y_{l - 1,1}}, \cdots {y_{l - 1,{N_{l,i}}}})^T}$ is the previous layer output, ${{\boldsymbol{\bf{b}}}_l} = {({b_{l,1}}, \cdots ,{b_{l,{N_{l,o}}}})^T}$, and ${{\boldsymbol{\bf{W}}}_l} = ({{\boldsymbol{\bf{w}}}_{l,1}}, \cdots ,{{\boldsymbol{\bf{w}}}_{l,{N_{l,o}}}})$, where ${{\boldsymbol{\bf{w}}}_{l,k}} = {({w_{l,k,1}}, \cdots ,{w_{l,k,{N_{l,i}}}})^T},k = 1, \cdots ,{N_{l,o}}$. As can be seen from the graph of \textit{Tanh} function, when result of ${{\boldsymbol{\bf{W}}}_l}^T{{\boldsymbol{\bf{y}}}_{l - 1}} + {{\boldsymbol{\bf{b}}}_l}$ is too small or too large, output ${\boldsymbol{\bf{y}}}_l$ will become saturated and the gradient will approach zero. This makes the output useless for the next layers and causes the gradient vanishing problem at the training phase. Variable initialization and batch normalization are proposed to deal with this problem.

To prevent signals flowing in the network from exploding to large values (including negative and positive), when passing a layer in forward propagation, variance of output values should be kept in the same as the inputs. That's to say, to choose proper initial values of weights and biases for each layer that satisfy $Var[{{\boldsymbol{\bf{y}}}_l}] = Var[{{\boldsymbol{\bf{y}}}_{l-1}}]$ is necessary. This is known as Xavier initialization\cite{glorot2010understanding}. Without considering biases and assuming activation function in the linear regime, relation of variance among output $\boldsymbol{\bf{y}}_l$, input $\boldsymbol{\bf{y}}_{l-1}$ and weight matrix ${\boldsymbol{\bf{W}}_l}$ is give by\cite{glorot2010understanding}:
\begin{equation} 
Var[{{\boldsymbol{\bf{y}}}_l}] = {N_{l,i}} \cdot Var[{{\boldsymbol{\bf{W}}}_l}] \cdot Var[{{\boldsymbol{\bf{y}}}_{l - 1}}]
\end{equation} 
Thus, ${N_{l,i}} \cdot Var[{{\boldsymbol{\bf{W}}}_l}] = 1$ is required to keep $Var[{{\boldsymbol{\bf{y}}}_l}] = Var[{{\boldsymbol{\bf{y}}}_{l-1}}]$. The similar argument can be made for the gradients when passing layers in backward propagation, that is to say, to make $Var[\partial L/\partial {{\boldsymbol{\bf{y}}}_l}] = Var[\partial L/\partial {{\boldsymbol{\bf{y}}}_{l - 1}}]$, ${N_{l,o}} \cdot Var[{{\boldsymbol{\bf{W}}}_l}] = 1$ is required. These two conditions can not both be satisfied at the same time, unless $N_{l,i}=N_{l,o}$. As \cite{glorot2010understanding} indicates, a compromise is given by:
\begin{equation} \label{EqWeightsInit}
Var[{{\boldsymbol{\bf{W}}}_l}] = \frac{2}{{{N_{l,i}} + {N_{l,o}}}}
\end{equation} 
For all the weights initialization in this paper, the above constraint is attached.

To further improve vanishing and exploding gradients problem, batch normalization technique\cite{wang2019batch} is applied in the networks in this paper. The purpose of batch normalization is to linearly transform layer's inputs to ones with zero means and unit variances, making them de-correlated and be kept in the active region of the activation functions, without corrupting the learned features.

The approach adopted in this paper is normalization via min-batch statistics. In SGD optimization training, the whole training data set is divided into min-batches to feed the network. Consider layer $l$ with input ${{\boldsymbol{\bf{y}}}_{l - 1}}$. As normalized values are directly feed into the activation functions, the normalizing should be taken place right before the non-linear activation. Therefore, denote ${{\boldsymbol{\bf{W}}}_l}^T{{\boldsymbol{\bf{y}}}_{l - 1}} + {{\boldsymbol{\bf{b}}}_l}$ as ${{\boldsymbol{\bf{z}}}_l} = {({z_{l,1}}, \cdots ,{z_{l,{N_{l,o}}}})^T}$, training min-batch of size $N_b$ as ${\mathcal B} = \{ {{\boldsymbol{\bf{z}}}_{l }}^{(1)}, \cdots ,{{\boldsymbol{\bf{z}}}_{l }}^{({N_b})}\} $. The batch normalization process includes the following steps:
\begin{enumerate}  

\item batch mean and variance computation: $\forall k \in \{ 1, \cdots ,{N_{l,o}}\}$
\begin{equation} 
\frac{1}{{{N_b}}}\sum\limits_{i = 1}^{{N_b}} {{z_{l,k}}^{(i)}}  \to {\mu _{{\mathcal B},k}}
\end{equation} 

\begin{equation} 
\frac{1}{{{N_b}}}\sum\limits_{i = 1}^{{N_b}} {{{\left( {{z_{l,k}}^{(i)} - {\mu _{{\mathcal B},k}}} \right)}^2}}  \to {\sigma _{{\mathcal B},k}}^2
\end{equation} 
For each dimension $k$ of every ${{\boldsymbol{\bf{z}}}_l}^{(i)}$ in batch ${\mathcal B}$, the calculated mean and variance ${\mathcal B}$ are ${\mu _{{\mathcal B},k}}$ and ${\sigma _{{\mathcal B},k}}^2$, respectively.

\item normalizing: $\forall k \in \{ 1, \cdots ,{N_{l,o}}\} , i \in \{ 1, \cdots ,{N_b}\}$
\begin{equation} 
\frac{{{z_{l,k}}^{(i)} - {\mu _{{\mathcal B},k}}}}{{\sqrt {{\sigma _{{\mathcal B},k}}^2 + \varepsilon } }} \to {{\hat z}_{l,k}}^{(i)}
\end{equation}
Every element $k$ of every ${{\boldsymbol{\bf{z}}}_l}^{(i)}$ in batch ${\mathcal B}$ is normalized to ${{\hat z}_{l,k}}^{(i)}$, where $\varepsilon$ is a constant added to the batch variance for numerical stability.

\item scaling and shifting: $\forall k \in \{ 1, \cdots ,{N_{l,o}}\} , i \in \{ 1, \cdots ,{N_b}\}$
\begin{equation} 
{\gamma _{l,k}}^{(i)} \cdot {{\hat z}_{l,k}}^{(i)} + {\beta _{l,k}}^{(i)} \to {a_{l,k}}^{(i)}
\end{equation} 
where ${\gamma _{l,k}}^{(i)}$ and ${\beta _{l,k}}^{(i)}$ are parameters to be learned along with the original weights and biases of the network during training. They are introduced to give the network chances to restore representation power when inputs are all normalized. The final batch normalized results are ${a_{l,k}}^{(i)},k = 1, \cdots ,{N_{l,o}},i = 1, \cdots ,{N_b}$, which then are fed into the activation functions producing layer outputs $\{ {{\boldsymbol{\bf{y}}}_l}^{(1)}, \cdots ,{{\boldsymbol{\bf{y}}}_l}^{({N_b})}\}$.

\end{enumerate}

In this paper, the process described above is added in every hidden layer in the SCMA decoding DNN and autoencoder. Experiments show that performances with and without batch normalization differ a lot. Batch normalization can largely improve the training speed and the decoding accuracy.

\section{Performance evaluation} \label{SecEva}

\subsection{Simulation setup}
Unless otherwise indicated, all the experiments conducted in this paper is through simulations with the configurations specified here. Consider SCMA system with $J=6$ users, $K=4$ sub-carriers and $M=4$, thus $m=2$, user codebooks given in \cite{asiapresentation} are used. The reason why we choose this codebook is twofold: 1) this is the only codebook explicitly given in public we can find; 2) the state of the art SCMA decoder from \cite{vameghestahbanati2017enabling} that we are going to compare performances with had used this codebook as well. In DL-SCMA, the number of nodes for input layer, hidden layers, and output layer is $2K=8$, $N_{HN}=48$, $2J=12$, respectively. The number of hidden layers $N_L$ is 6. The training set is composed of 5 groups, each contains 500000 samples and is randomly generated at $E_b/N_0=2,3,4,5,6$, respectively, thus the total number of training data is 2500000. For AE-SCMA, there are 4 and 5 hidden layers for the encoder and decoder part and the number of nodes is 32 and 48, respectively. The number of training data is 2000000, also randomly generated. In the training phase, $E_b/N_0$ is configured to 5. The gradient descent training optimizer ADAM with learning rate of 0.0001 is adopted. Table \ref{TabExpParameters} summarizes the experiment configurations.

All simulations are run on a DELL Precision Tower 7810 workstation with 2 Inter Xeon E5-2630V4@2.2GHz CPUs, each with 10 cores, 32GB RAM and an 8GB Nvidia Quadro K5200 GPU. Software environment includes Ubuntu 16.04 operating system, Tensorflow 1.6.0 and Python 2.7.12.

\begin{table}[htbp]
 \caption{\label{TabExpParameters} Summary of the experiment configurations.}
 \centering
 \begin{tabular}{ll}
  \toprule
    Parameter & Value \\
  \midrule
   Numb. of SCMA users $J$ & 6 \\
   Numb. of sub-carriers $K$ & 4 \\
   Numb. of codewords for each user $M$ & 4 \\
   
   Numb. of nodes for input layer & 8 \\
   Numb. of nodes for output layer & 12 \\
   
   Numb. of nodes for DL-SCMA's hidden layer & 48 \\
   Numb. of nodes for AE-SCMA's encoder part & 32 \\
   Numb. of nodes for AE-SCMA's decoder part & 48 \\
   
   Numb. of hidden layers for DL-SCMA & 6 \\
   Numb. of hidden layers for AE-SCMA's encoder part & 4 \\
   Numb. of hidden layers for AE-SCMA's decoder part & 5 \\
   
   Numb. of training samples for DL-SCMA & 500000 \\
   Training $E_b/N_0$ for DL-SCMA & 2, 3, 4, 5, 6 \\
   
   Numb. of training samples for AE-SCMA & 200000 \\
   Training $E_b/N_0$ for AE-SCMA & 5 \\
   
   Training optimizer & ADAM \\
   Learning rate & 0.0001 \\
   
  \bottomrule
 \end{tabular}
\end{table}

\subsection{Decoding accuracy}

We mainly compare our proposed schemes with the traditional SCMA decoding algorithm MPA\cite{nikopour2013sparse} but computed in logarithm domain, here called Log-MPA. The state of the art non-DL SCMA detection scheme called modified sphere decoding (MSD) from \cite{vameghestahbanati2017enabling} is also taken into account. Moreover, two schemes called D-SCMA+DNN and D-SCMA+MPA from the related work\cite{kim2018deep}, which are also based on DL methods, are on the comparison list as well. The idea of D-SCMA+DNN is similar to our proposed autoencoder, but the structures are different. In the encoder part, D-SCMA+DNN is composed of a branch of fully connected (FC) networks, each of them is corresponding to an edge of the SCMA bipartite graph. While our approach is constructed by stacking $J$ FC networks and $J$ sparse mapping binary vectors $s_j$ together for the $J$ users to form the SCMA coding process. Our approach can easily generalize SCMA, by setting large fractions of values in $s_j$ to 1, the DCMA is proposed.

As described in section \ref{SecModel}, in SCMA, transmitted bits are grouped into sets and then mapped to codewords that are actually related to transmitting signals, we call these grouped bits symbols and define symbol error rate (SER) as the percentage of decoded grouped bits different from the transmitting grouped bits. SER and BER are related, but not the same. SER is actually more practical than BER in a real communication application. 

\begin{figure}
  \centering
    \includegraphics[width=0.8\textwidth]{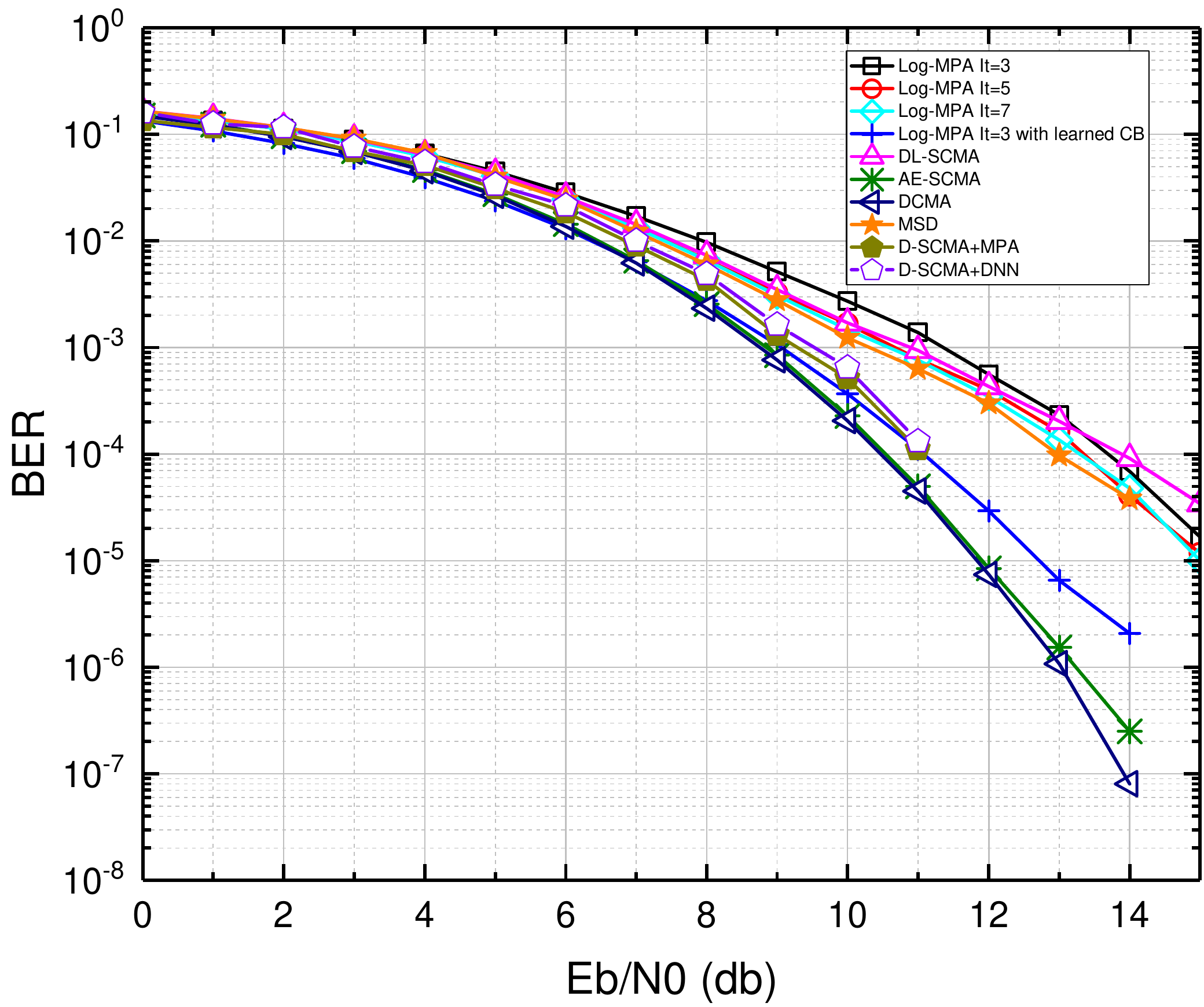}
  \caption{BER performance comparisons.}
  \label{FigBERPER}
\end{figure}

Fig. \ref{FigBERPER} demonstrates the BER comparing results. When $3< E_b/N_0 < 14$, DL-SCMA is superior to Log-MPA with 3 iterations, and close to 5 iterations. When $E_b/N_0$ is small enough, DNN is hard to learn the causal relationship between input and output, and when $E_b/N_0$ is large enough, the learned network seems too conservative. Both of these cases weaken DL-SCMA's ability to generalize SCMA decoding process. Compare to MSD scheme, DL-SCMA has equivalent BERs when at small $E_b/N_0$, but has larger BERs when at large $E_b/N_0$. It is notable that D-SCMA+DNN and D-SCMA+MPA use  codebooks generated by another DNN encoder\cite{kim2018deep}, which are different from the ones our DL-SCMA and MSD used. Due to this, D-SCMA+DNN and D-SCMA+MPA have better BER performances than DL-SCMA. For a fair comparison, D-SCMA+DNN and D-SCMA+MPA should be compared to our AE-SCMA scheme and the Log-MPA with learned codebook, respectively.

In all situations, AE-SCMA performs better than Log-MPA, MSD, D-SCMA+DNN and D-SCMA+MPA significantly. This is surely due to the optimal learned codebooks. It also shows that DCMA slightly outperforms SCMA. This is reasonable for that DCMA makes more use of sub-carriers thus gives more information of the transmitting bits for decoding. Section \ref{SecDis} gives more discussions why DCMA could improve the performance. In addition, to further evaluate the learned codebooks, we have conducted the Log-MPA algorithm using these codebooks, whose result is presented by the '+' symbol denoted line. It can be seen that Log-MPA with learned codebooks indeed surpasses the one with original codebooks, as well as MSD. Moreover, it also surpasses AE-SCMA when $E_b/N_0$ is less than 5. The reason for this achievement is similar to the analysis of the comparison between DL-SCMA and Log-MPA when in small $E_b/N_0$ range.

\begin{figure*}
	\centering
	\begin{subfigure}{0.45\textwidth}
		\includegraphics[width=\textwidth]{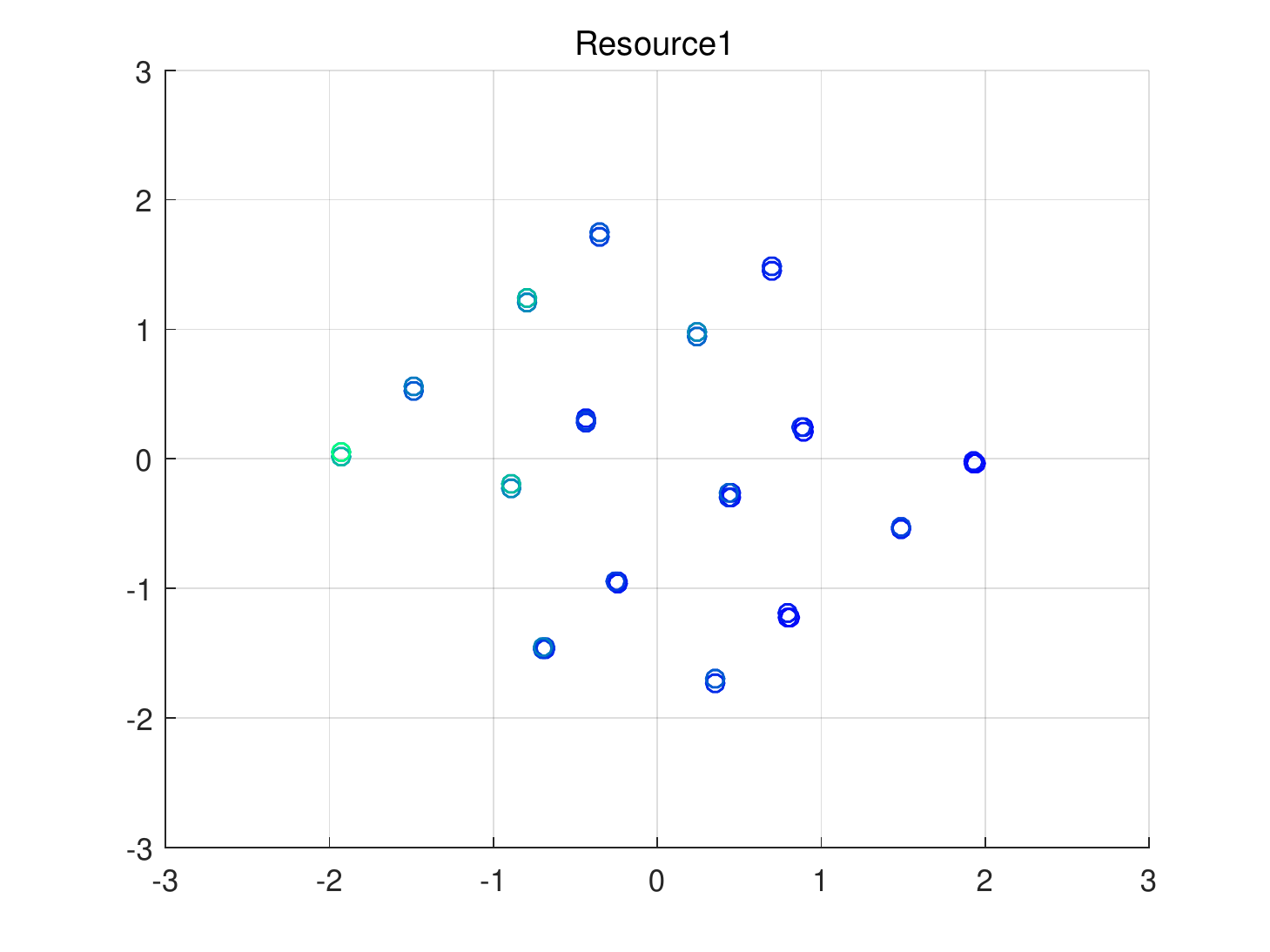}
		\caption{ }
	\end{subfigure}
	\begin{subfigure}{0.45\textwidth}
		\includegraphics[width=\textwidth]{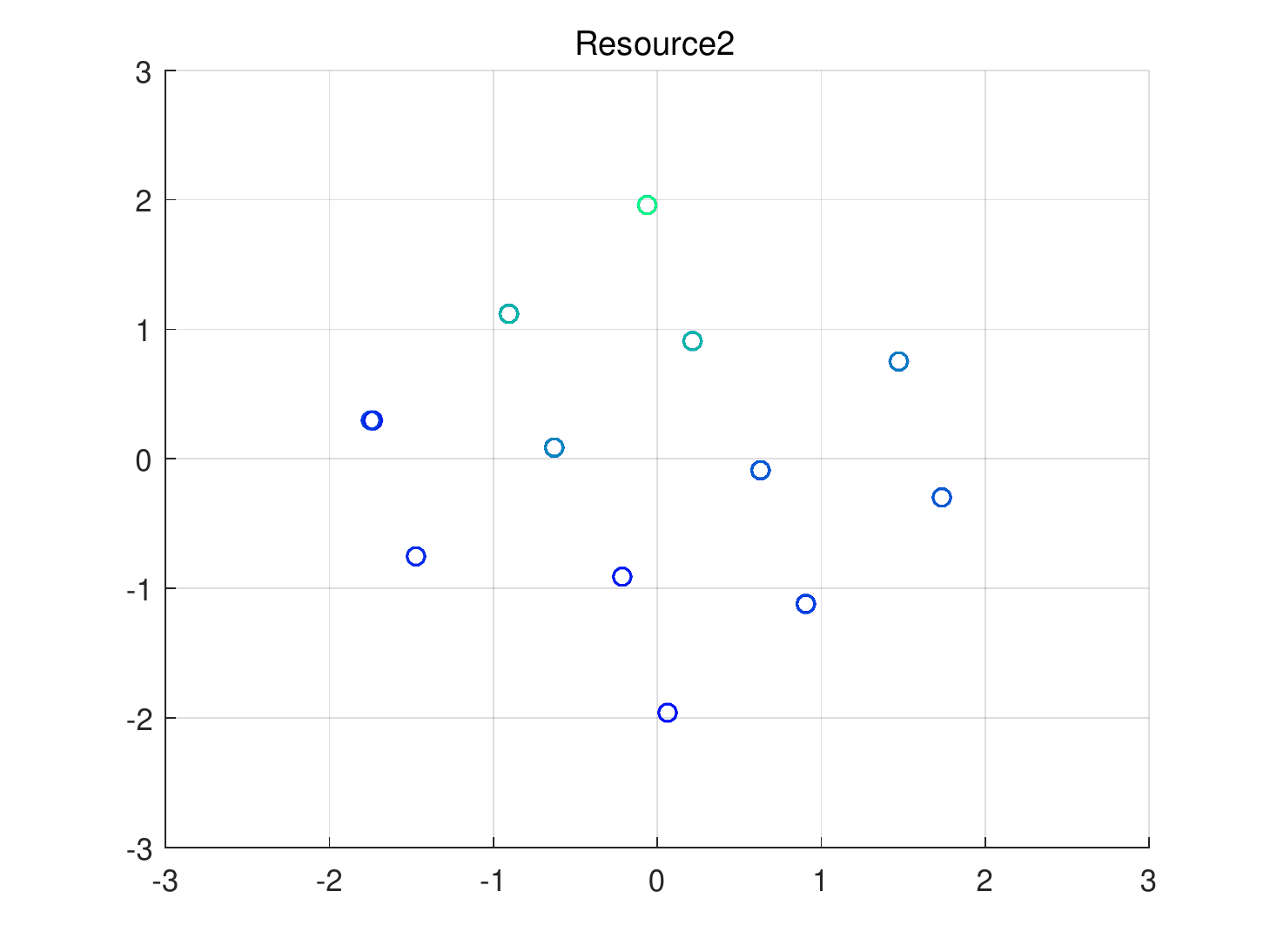}
		\caption{ }
	\end{subfigure}
	\begin{subfigure}{0.45\textwidth}
		\includegraphics[width=\textwidth]{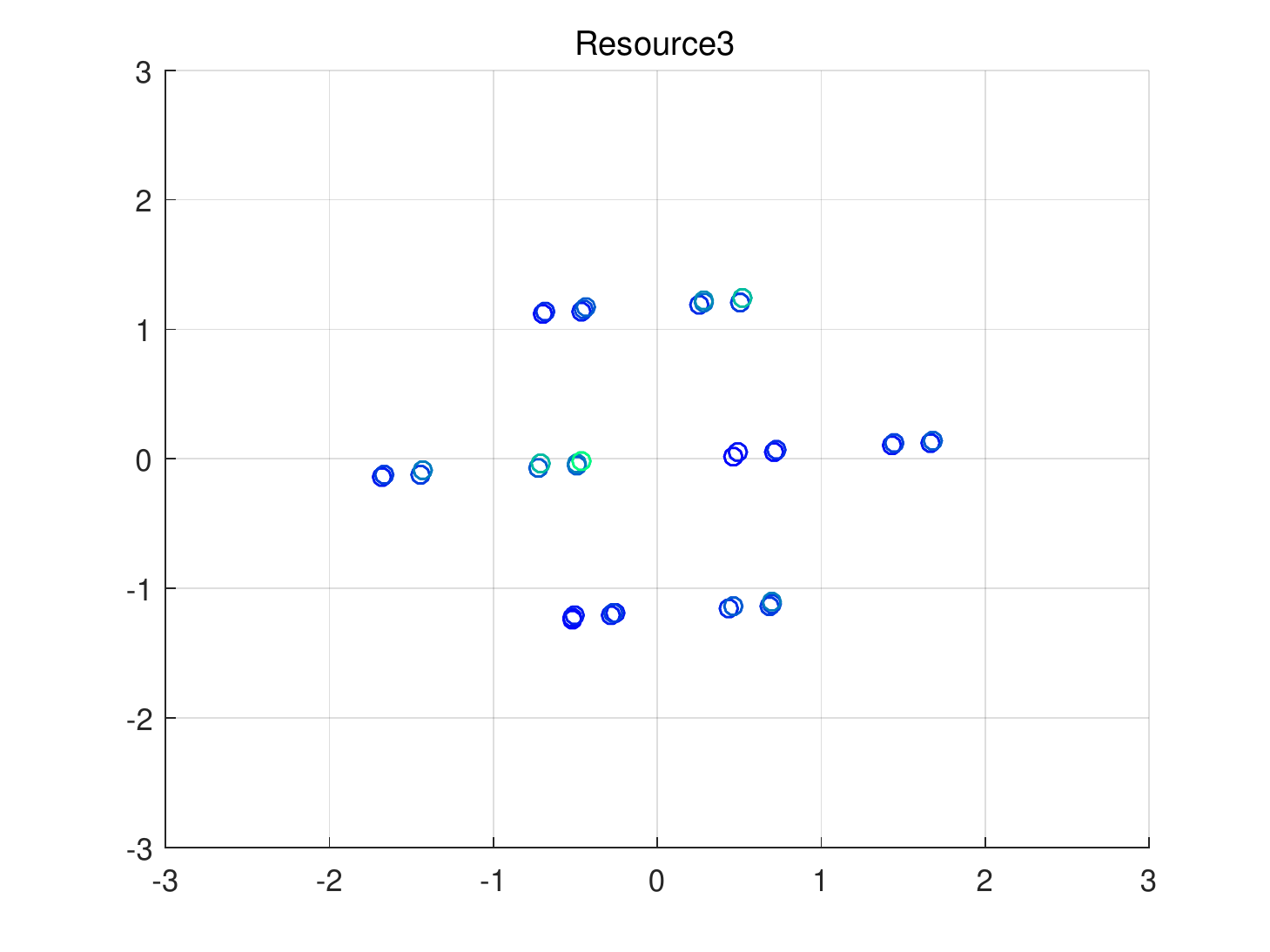}
		\caption{ }
	\end{subfigure}
	\begin{subfigure}{0.45\textwidth}
		\includegraphics[width=\textwidth]{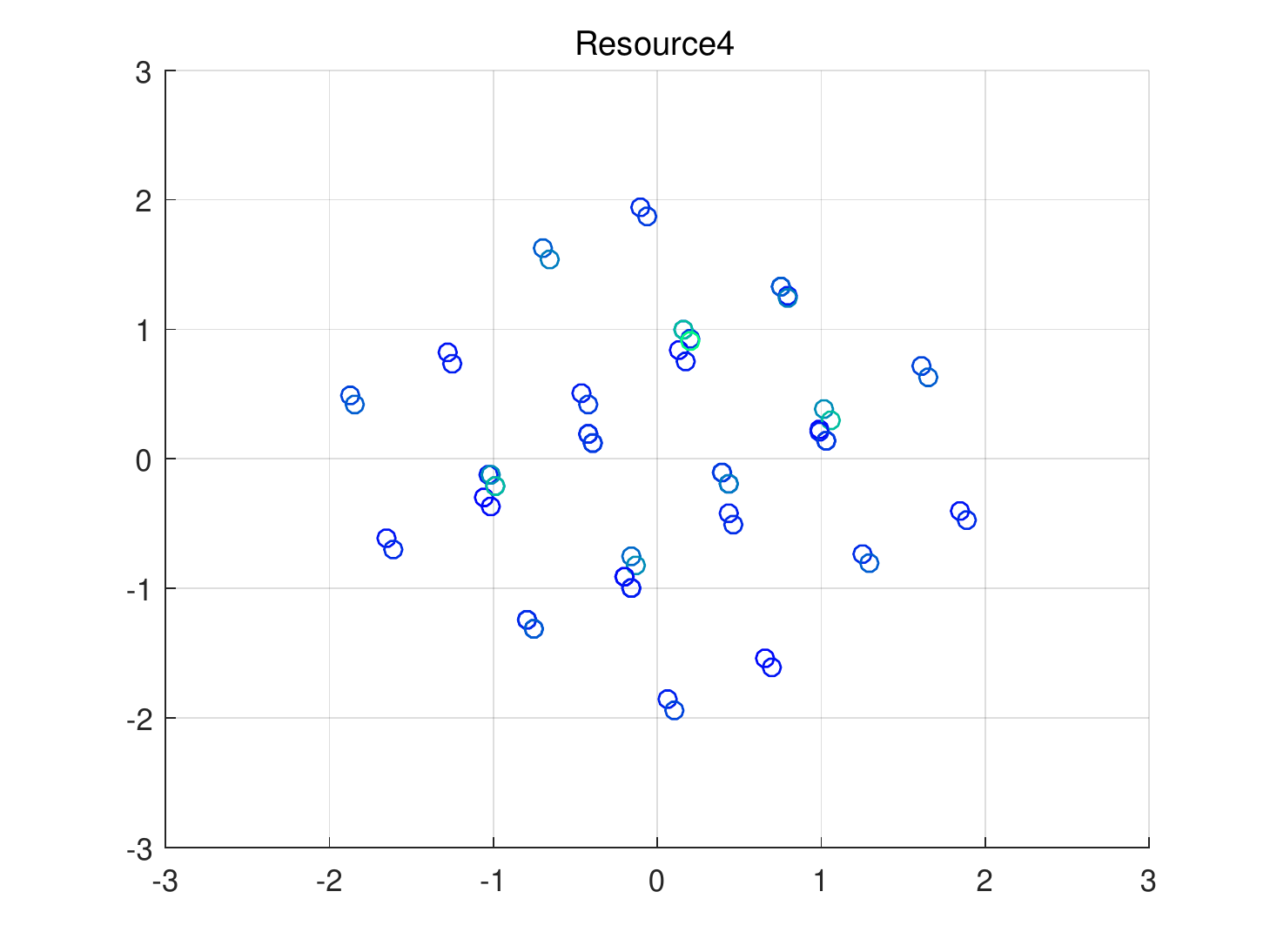}
		\caption{ }
	\end{subfigure}
	\caption{The projections of the learned codebook constellations over (a)RE1, (b)RE2, (c)RE3 and (d)RE4.}
	\label{FigCB}
\end{figure*}

Fig. \ref{FigCB} depicts the projections of the learned codebooks' superposition constellations over the 4 resources(sub-carriers). As can be seen, there are certain patterns hidden in them. Unlike constellation points of codebook from \cite{asiapresentation}, which are spread apart as much as possible, points are grouped together in some form of patterns. We believe that these patterns are related to the SCMA factor graph and the symbols mapping. Due to this points grouping, Euclidean distances between groups are enlarged, thus, for AWGN, group decoding error ratio reduces. As mapping symbols are transmitted over several(here is two) resources, as long as symbols can uniquely be determined by the intersection of groups from different resources, constellation points grouping is reasonable as well as feasible. By investigating the learned codebooks, we found that AE-SCMA constructed codewords exactly complied with this idea. From (d) of Fig. \ref{FigCB}, it can be seen that points grouping is a little imperfect. We believe that this is due to that AE-SCMA did not converged to the optimal solution during training, which means that there is still space for improvement in obtaining better codebooks.

\begin{figure*}
	\centering
	\begin{subfigure}{0.45\textwidth}
		\includegraphics[width=\textwidth]{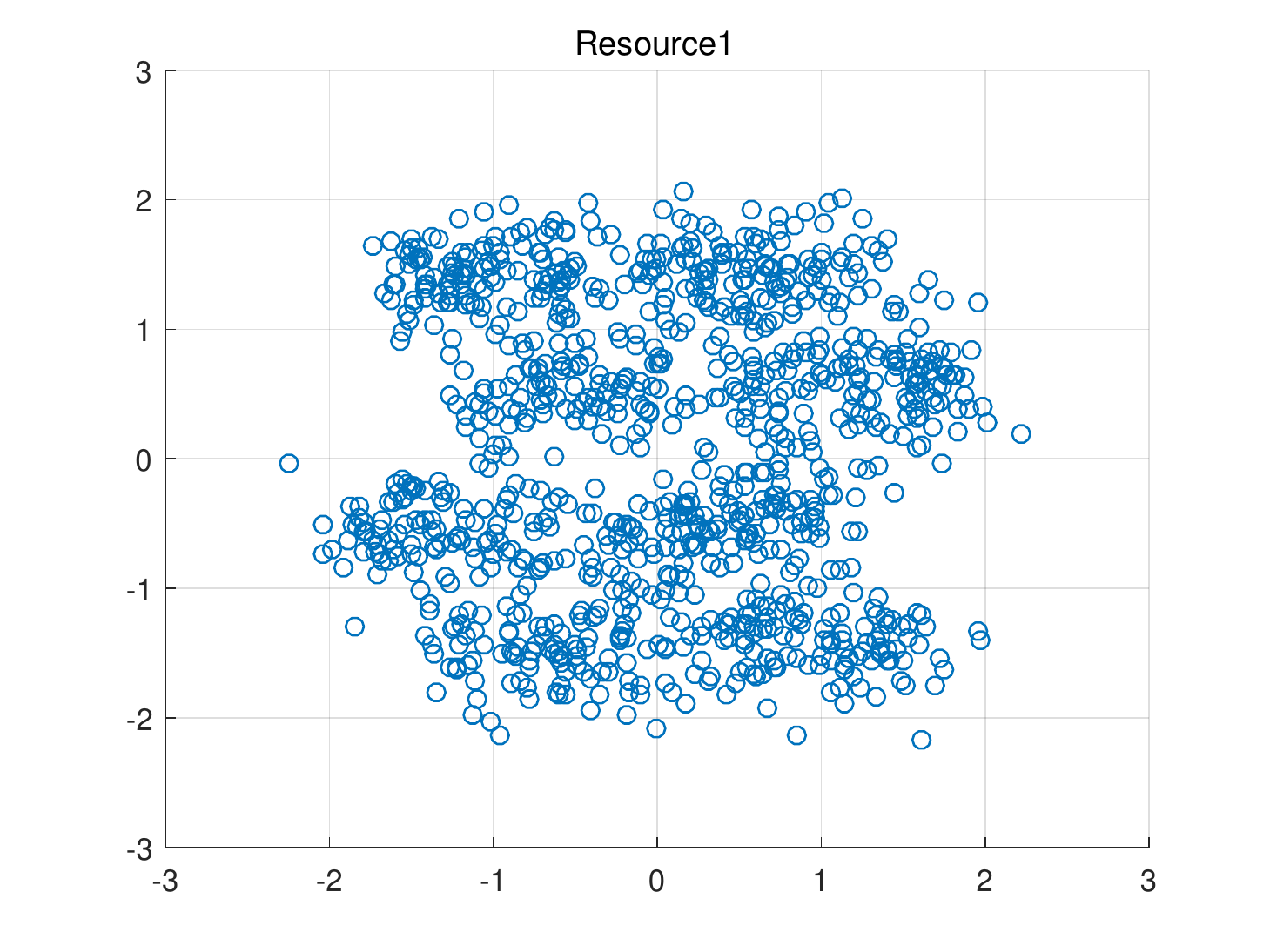}
		\caption{ }
	\end{subfigure}
	\begin{subfigure}{0.45\textwidth}
		\includegraphics[width=\textwidth]{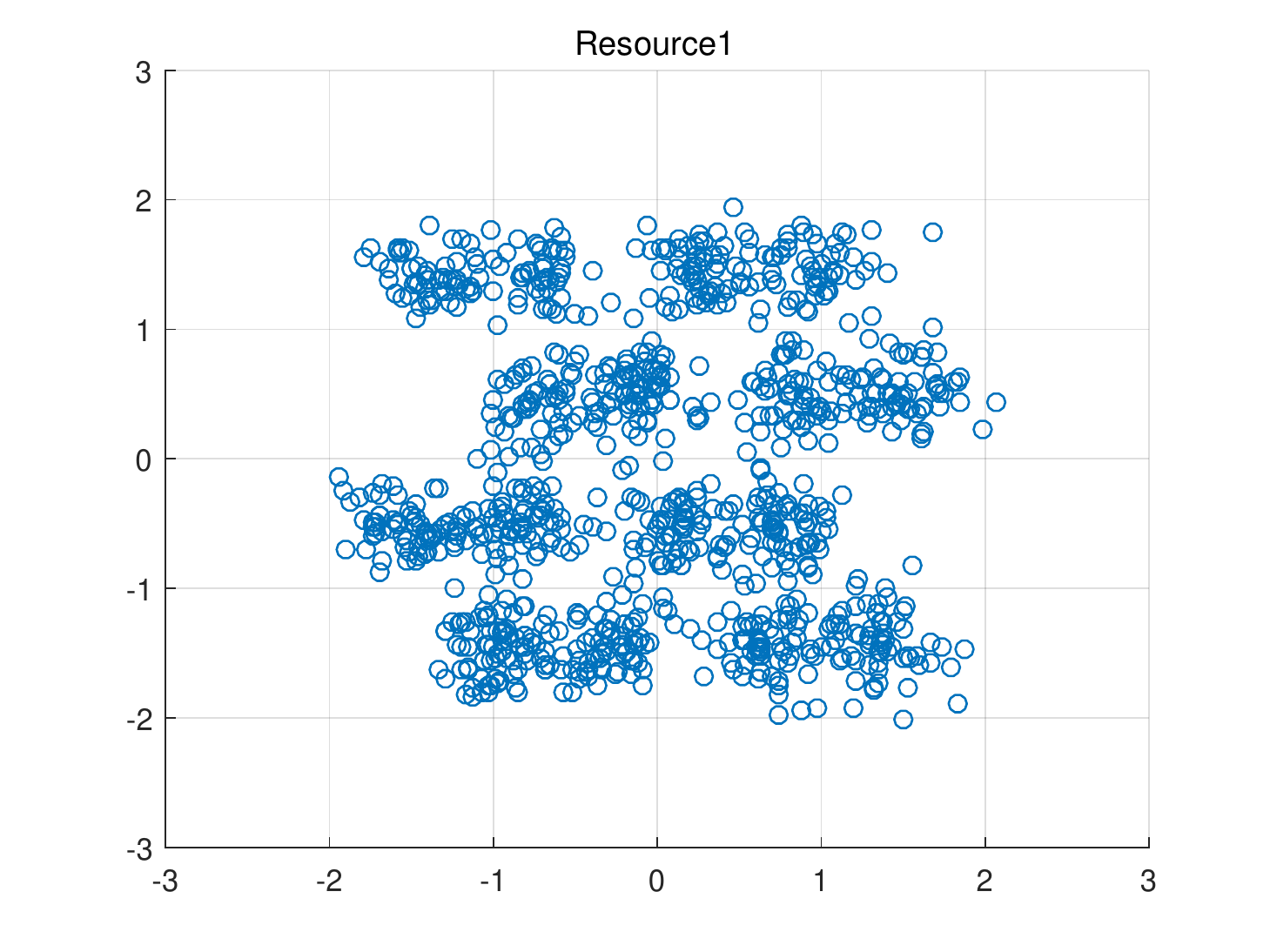}
		\caption{ }
	\end{subfigure}
	\begin{subfigure}{0.45\textwidth}
		\includegraphics[width=\textwidth]{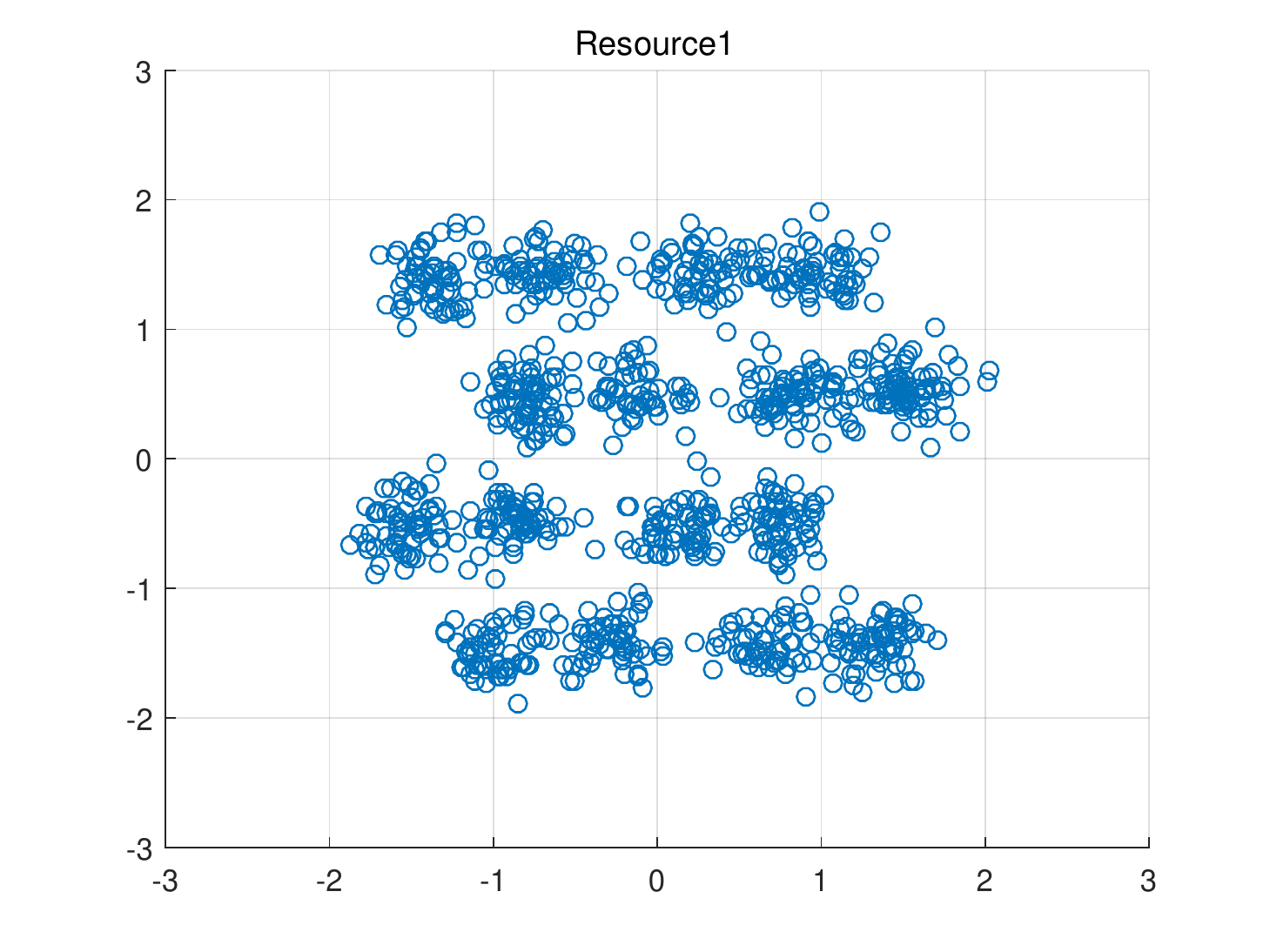}
		\caption{ }
	\end{subfigure}
	\begin{subfigure}{0.45\textwidth}
		\includegraphics[width=\textwidth]{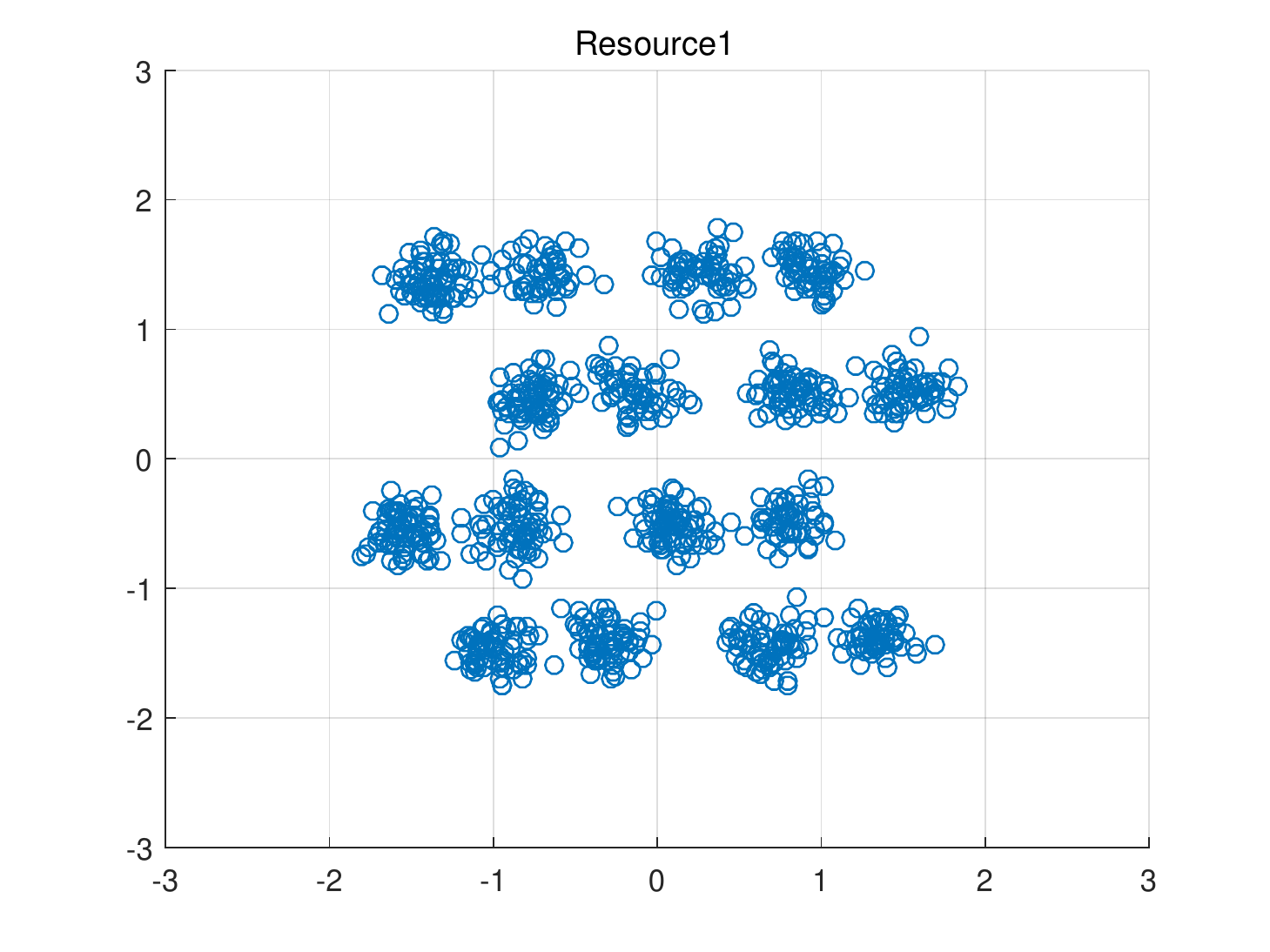}
		\caption{ }
	\end{subfigure}
	\caption{Constellation diagrams of the received signals over RE1 with different SNRs when using the learned codebooks. (a)$E_b/N_0=4$, (b)$E_b/N_0=6$, (c)$E_b/N_0=8$ and (d)$E_b/N_0=10$.}
	\label{FigSigConstelAEG}
\end{figure*}

\begin{figure*}
	\centering
	\begin{subfigure}{0.45\textwidth}
		\includegraphics[width=\textwidth]{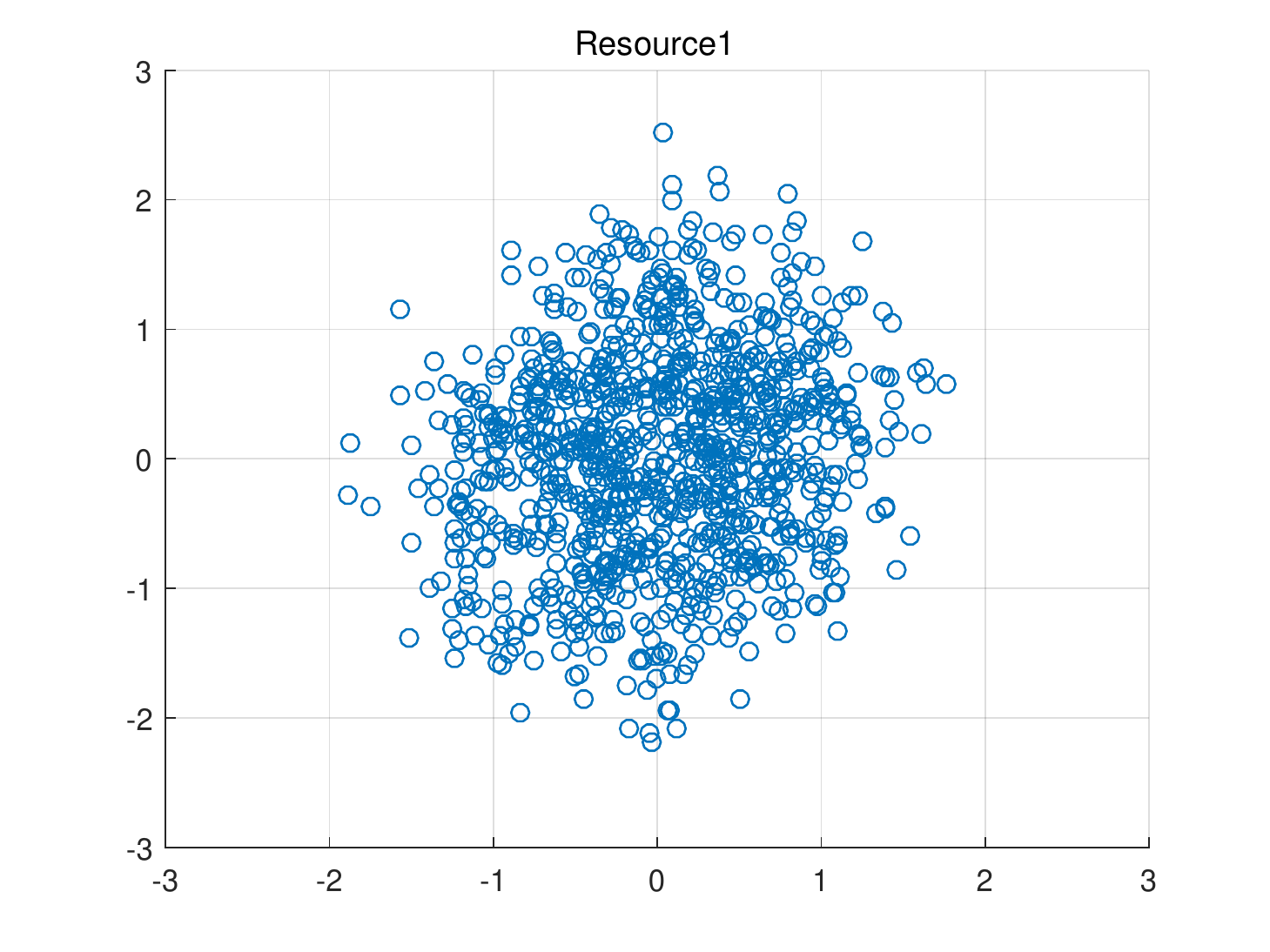}
		\caption{ }
	\end{subfigure}
	\begin{subfigure}{0.45\textwidth}
		\includegraphics[width=\textwidth]{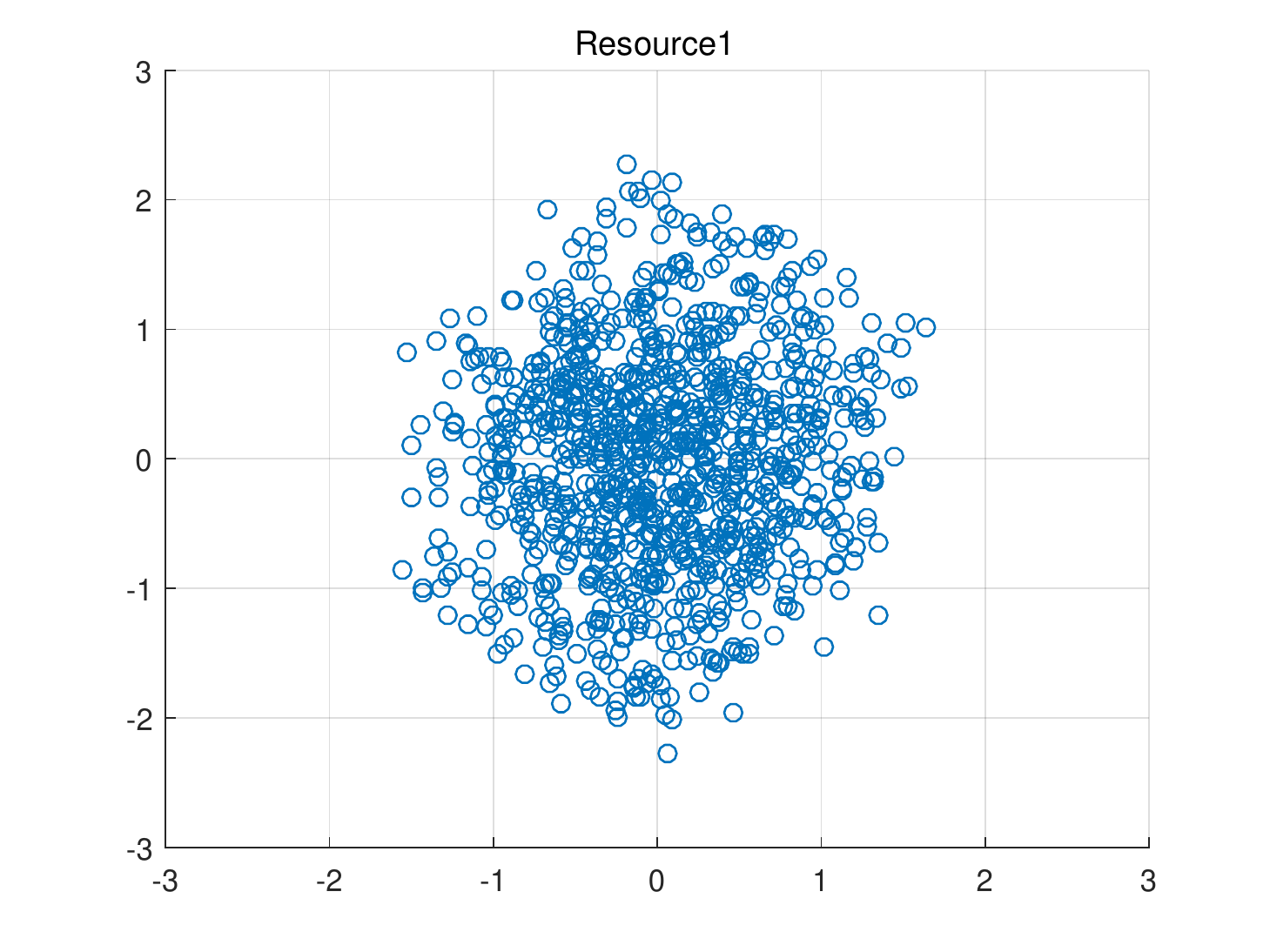}
		\caption{ }
	\end{subfigure}
	\begin{subfigure}{0.45\textwidth}
		\includegraphics[width=\textwidth]{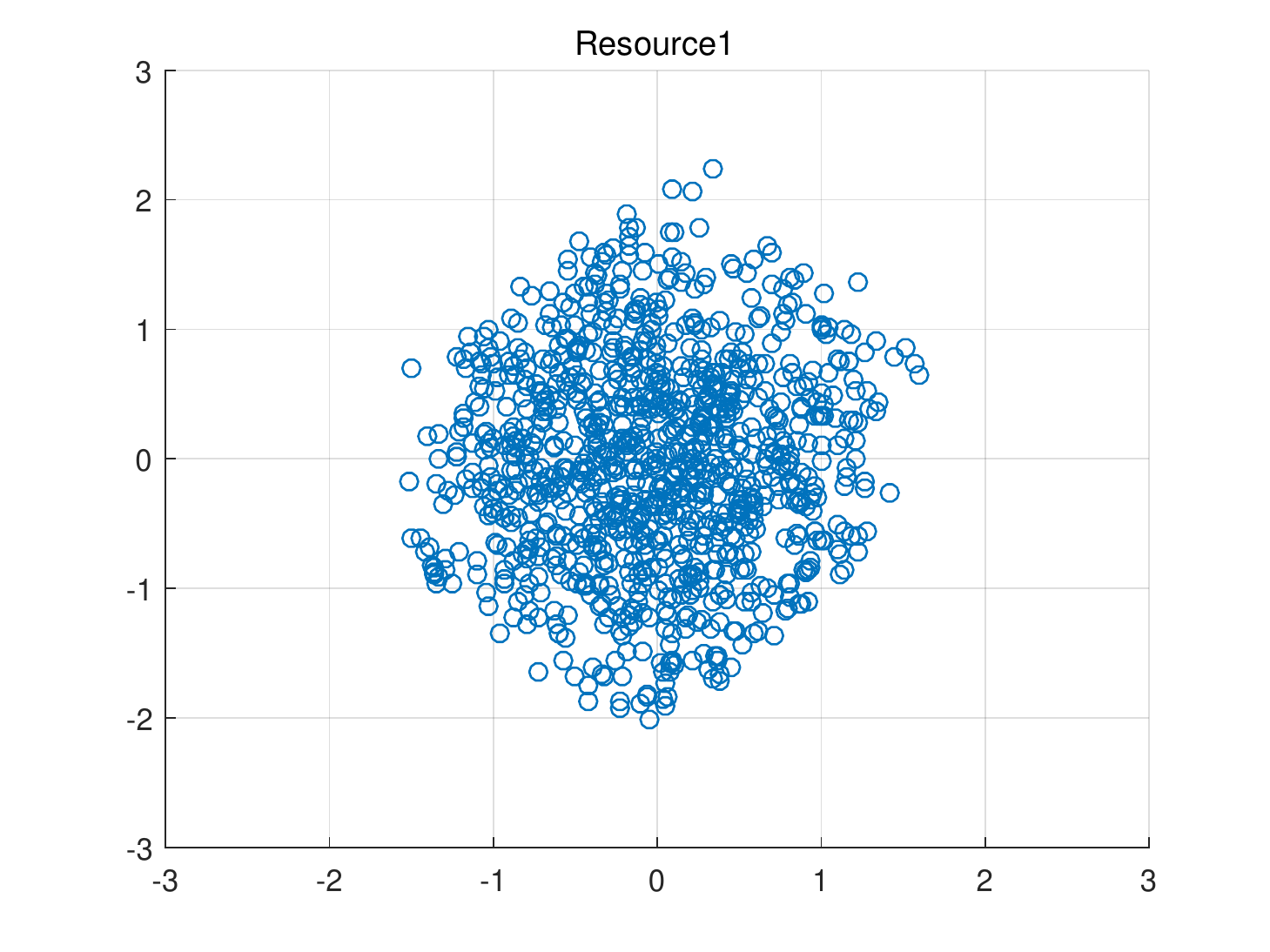}
		\caption{ }
	\end{subfigure}
	\begin{subfigure}{0.45\textwidth}
		\includegraphics[width=\textwidth]{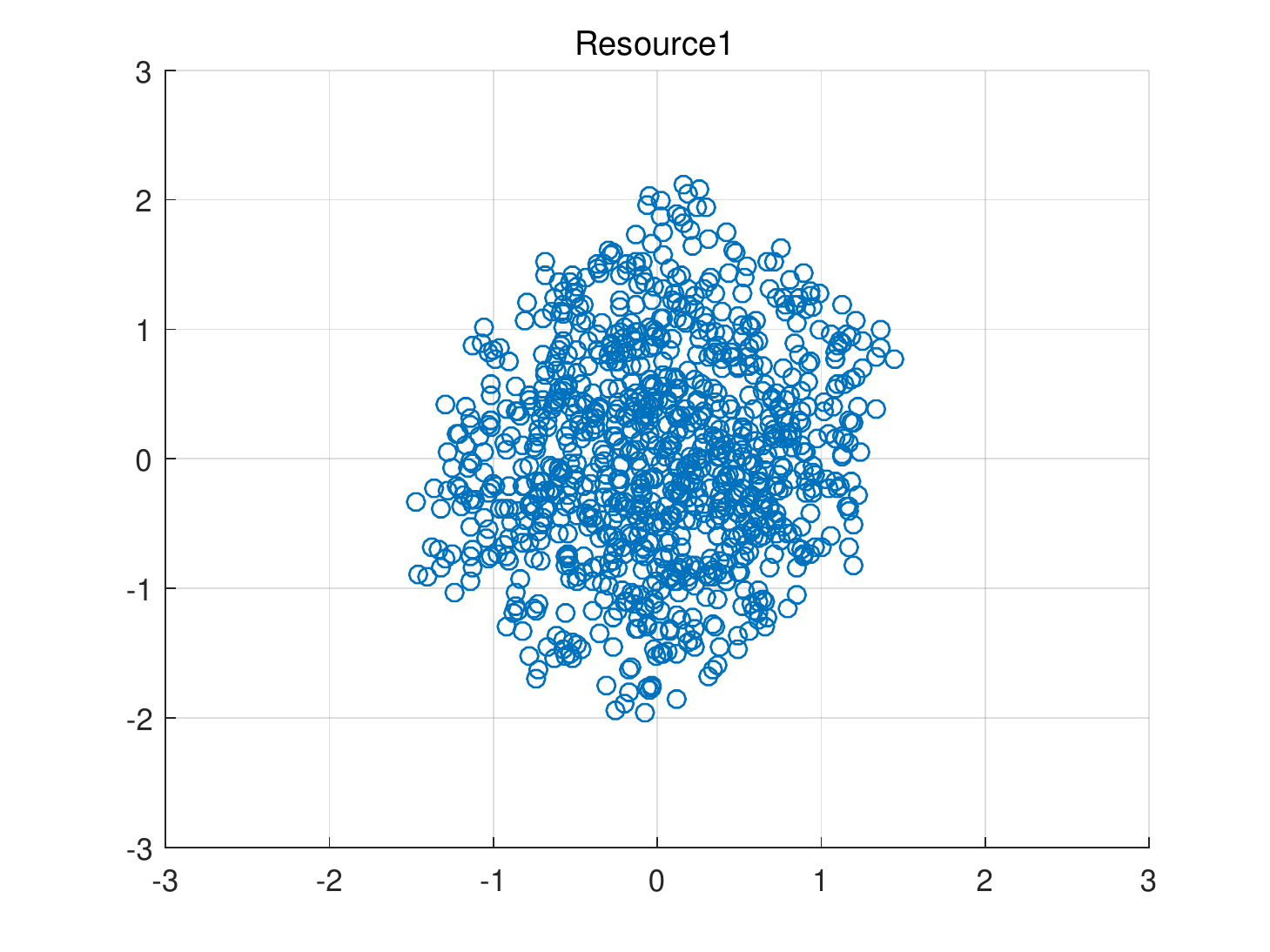}
		\caption{ }
	\end{subfigure}
	\caption{Constellation diagrams of the received signals over RE1 with different SNRs when using the codebook in \cite{asiapresentation}. (a)$E_b/N_0=4$, (b)$E_b/N_0=6$, (c)$E_b/N_0=8$ and (d)$E_b/N_0=10$.}
	\label{FigSigConstelTRD}
\end{figure*}

Fig. \ref{FigSigConstelAEG} and \ref{FigSigConstelTRD} are constellation diagrams of the received signals with different SNRs when using the learned codebooks and the codebooks from \cite{asiapresentation} respectively. Due to space limitations, only resource 1's signals are given. Other resources have the same situations, thus are ignored here. As analyzed above, points grouping causing larger Euclidean distances, different groups of constellation points of the learned codebooks maintain better differences in position than the codebooks from \cite{asiapresentation} when facing channel noises, therefore resist and distinguish more severely corrupted signals, so as to reduce the decoding error and enhance the whole decoder's performance. The formation shape of the constellations in fig. \ref{FigSigConstelAEG} is consistent with fig. \ref{FigCB}.(a), although certain angular rotation is emerged, which is the result of channel coefficients adding in.

\begin{figure}
  \centering
    \includegraphics[width=0.8\textwidth]{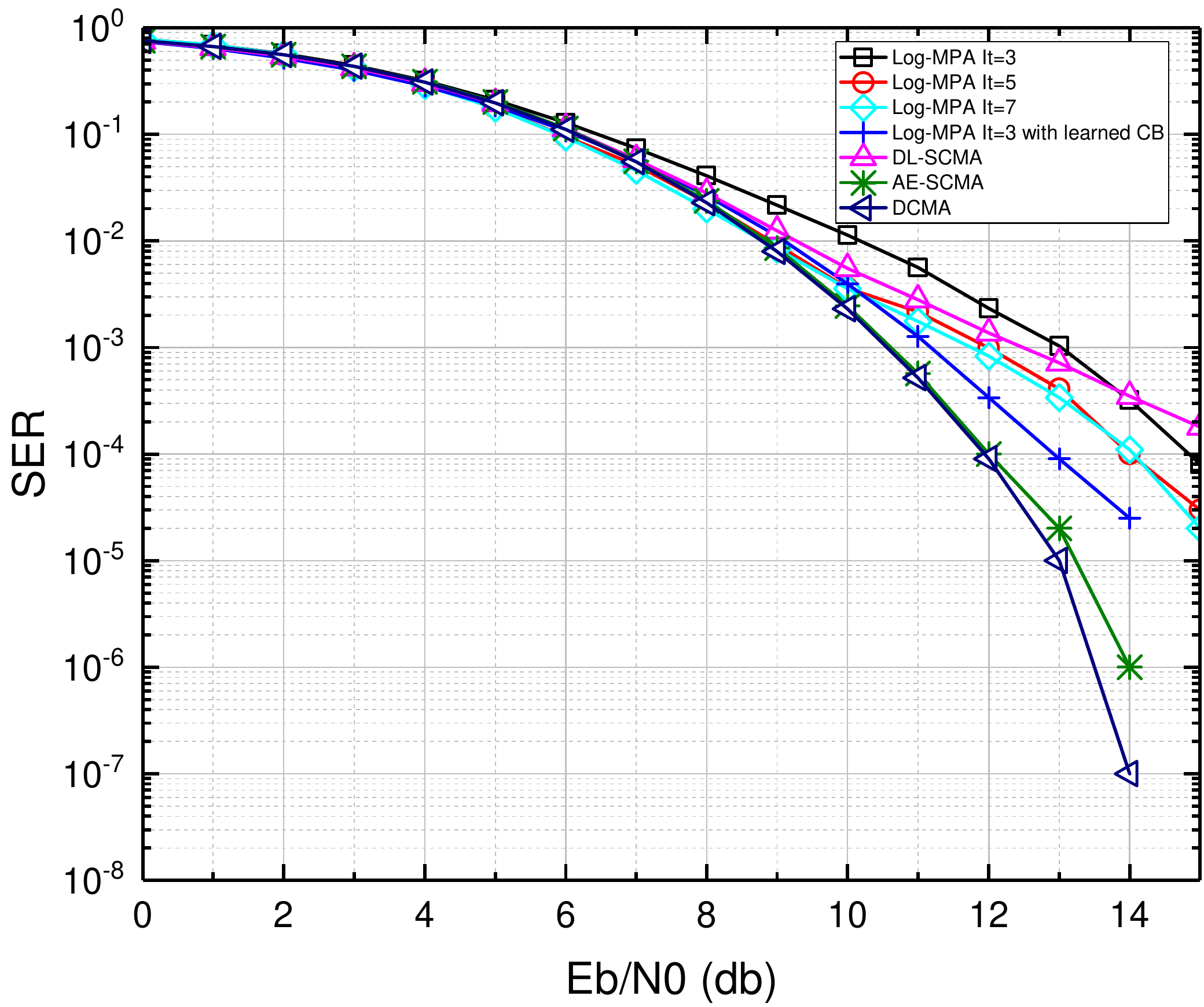}
  \caption{SER performance comparisons.}
  \label{FigSERPER}
\end{figure}

Fig. \ref{FigSERPER} shows the SER performances. As we've pointed out previously, SER is related to BER, thus similar results are expected. The difference lies in that when $E_b/N_0>14$, including $E_b/N_0>3$, DL-SCMA is better than Log-MPA with 3 iterations. This means that DL-SCMA has learned the exact mapping relationships between the received interweave signals and the corresponding codewords (grouped bits, or symbols), instead of individual bit streams. That's to say, DL-SCMA has high probability of that it decodes the bits of a group either all of them right or most of them wrong. When $E_b/N_0$ is small, this leads to obvious different results. AE-SCMA still has better performance in SER, especially when $E_b/N_0$ is large.

\begin{figure}
  \centering
    \includegraphics[width=0.8\textwidth]{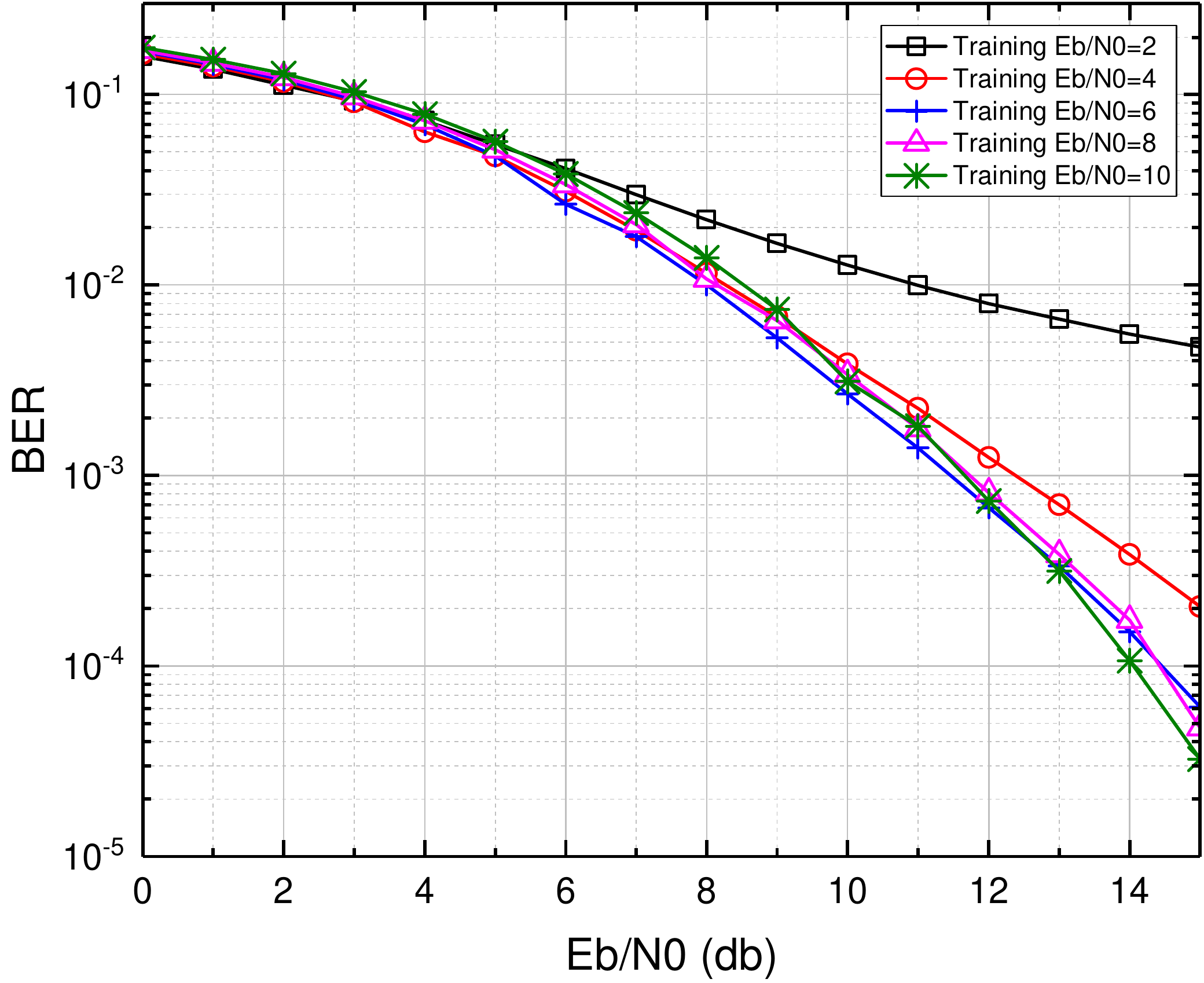}
  \caption{BER performances of DL-SCMA trained by various $E_b/N_0$.}
  \label{FigBERTrainingEbNo}
\end{figure}

\begin{figure}
  \centering
    \includegraphics[width=0.8\textwidth]{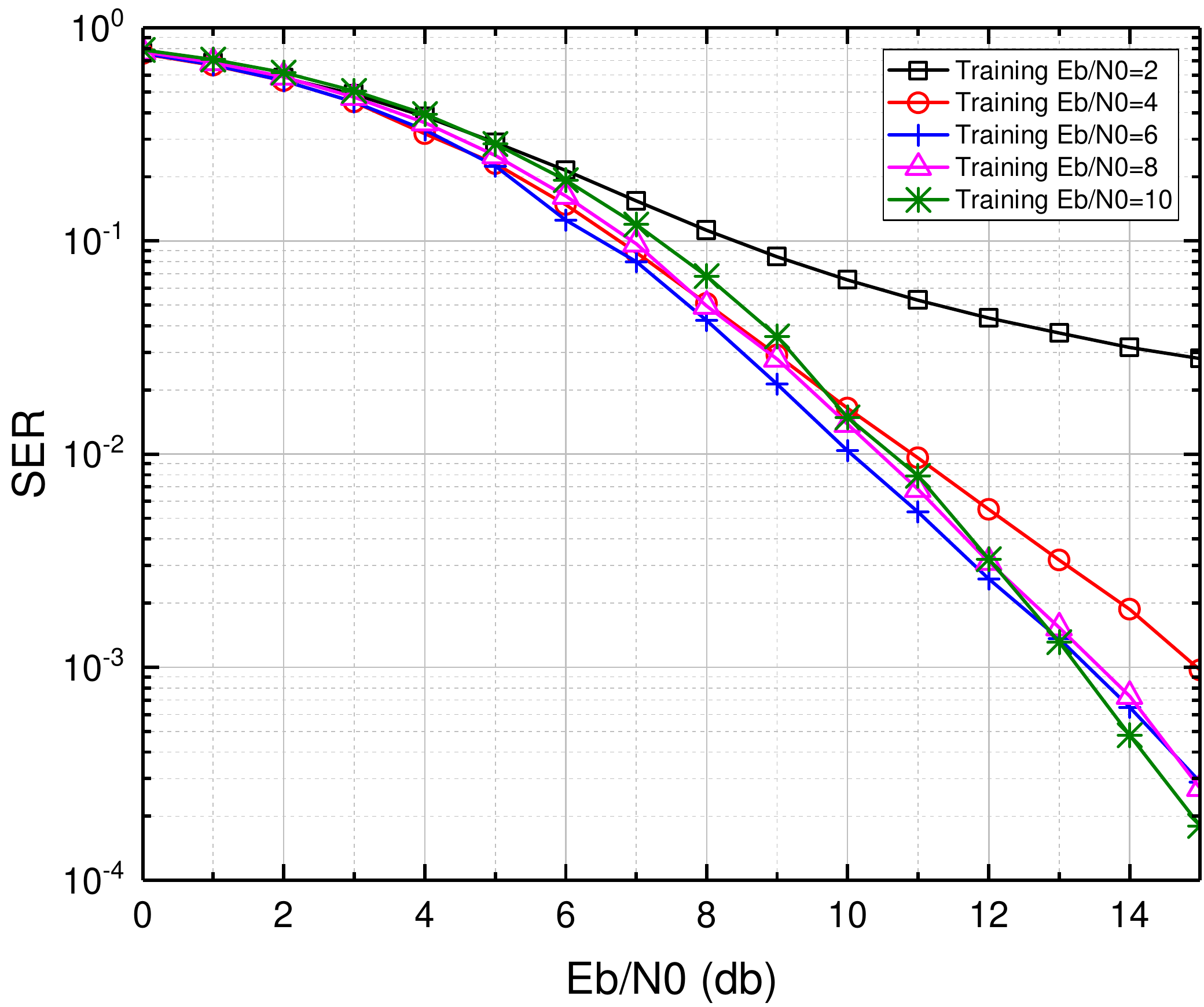}
  \caption{SER performances of DL-SCMA trained by various $E_b/N_0$.}
  \label{FigSERTrainingEbNo}
\end{figure}

To investigate how training samples generated by different $E_b/N_0$ levels can affect DL-SCMA's performance, comprehensive experiments have also been implemented in this paper. Here, we pick out 5 typical cases where $E_b/N_0=2,4,6,8,10$ for comparison. As shown in Fig. \ref{FigBERTrainingEbNo} and \ref{FigSERTrainingEbNo}, on average, the best $E_b/N_0$ level for generating training set is 6, which lies in about the middle of the range of $[0,10]$, not too low nor too high. It conforms to the previous analysis in section \ref{SecSch}. Generally, low $E_b/N_0$ trained networks perform well when decoding low $E_b/N_0$ testing data sets, but gradually decay when testing $E_b/N_0$ grows. The networks trained by a high $E_b/N_0$ have opposite situations.
It also can be seen that the decoder trained by $E_b/N_0$ 6 has superior performance to those trained by 8 and 10 in most testing $E_b/N_0$ cases (including 8 and 10). This is because training dataset in proper $E_b/N_0$ (here is 6) contains proper structures suitable for the DNN training process to extract the real hidden features related to essences a correct decoder should have. Although decoders trained by 8 and 10 $E_b/N_0$ may have better accuracy in decoding their training datasets, when facing to a testing dataset, they show inferior performances to the decoder trained by 6. That's to say the decoder trained by proper $E_b/N_0$ has better generalization ability. 
All of these prove that DL-SCMA and AE-SCMA can construct appropriate neural networks for SCMA decoding by learning from noise-corrupted SCMA signals.


\subsection{Computational complexity}
Log-MPA's complexity is ${\mathcal O}\left( {{M^{{d_f}}}} \right)$, where $d_f$ is the overlapping degree as defined previously. While the complexity of a DNN is ${\mathcal O}\left( {{N_{L}} \cdot {N_{HN}}^2} \right)$. Specifically, the average complexities in terms of multiplication, addition and log/exp operations of Log-MPA and DL-SCMA are provided in Table \ref{TabComplexity}\cite{vameghestahbanati2017enabling}, where $I_t$ is the number of iterations. For a fair comparison, we have normalized the different arithmetic operations based on the complexity of arithmetic operations\cite{brent2010modern}. Specifically, we normalize an addition to one unit of complexity, a multiplication to 10 units, and an exponentiation to 20 units\cite{chen2018joint}. As a result of the parameters configured in the simulation, Table \ref{TabNormedComlexity}, gives the specific number of the three arithmetic operations and the summed normalized complexities for the Log-MPA and DL-SCMA. It can be observed that DL-SCMA reduces $5.6\%$, $23.3\%$ and $35.4\%$ complexities compare to Log-MPA at 3, 5 and 7 iterations, respectively. As indicated in the previous subsection, DL-SCMA shows inferior performance compare to Log-MPA with iteration larger than 3. However, considering the computational complexity combined decoding accuracy, this is worthy.

The complexity of MSD depends on the number of symbols and the average number of visited layers\cite{vameghestahbanati2017enabling} when searching the tree, which is inversely proportional to the SNR. For this specific case, there are totally $4^6=4096$ symbols. According to our estimation, when $E_b/N_0$ is lower than 2, DL-SCMA has less complexity than MSD. Computational complexities of D-SCMA+MPA and D-SCMA+DNN have the same order of Log-MPA and DL-SCMA, respectively. But D-SCMA+DNN's neural network has much more hidden nodes compare to DL-SCMA and AE-SCMA. For this specific simulation, there are 512 hidden nodes for each hidden layer in D-SCMA+DNN, while DL-SCMA has only 48. There are 6 hidden layers each with 32 hidden nodes in the network in \cite{kim2018deep} for generating learned codebooks used by D-SCMA+MPA and D-SCMA+DNN, while the corresponding encoder part of AE-SCMA has 4 layers each also with 32 nodes. Therefore, in terms of computation, D-SCMA+DNN and its codebook generating DNN are more complex than the proposed DL-SCMA and AE-SCMA, respectively.

\begin{table*}[htbp]
\newcommand{\tabincell}[2]{\begin{tabular}{@{}#1@{}}#2\end{tabular}} 
 \caption{\label{TabComplexity} Average complexities of Log-MPA and DL-SCMA}
 \centering
 \begin{tabular}{ccc}
  \toprule
    & Log-MPA & \tabincell{c}{DL-SCMA}\\
  \midrule
  Multiplication & $MK{d_f}\left( {4{d_f}{M^{{d_f} - 1}} + 5} \right)$ &  ${N_{HN}}\left( {2K + {N_L}{N_{HN}} + 2J} \right)$\\
  Addition & $\begin{array}{l}
MK{d_f}({M^{{d_f} - 1}}(4{d_f} - 2
 + {I_t}(2 + \frac{1}{M})) + {I_t}(2 - \frac{1}{N}) + 5)
\end{array}$ & ${N_{HN}}\left( {{N_L} - 1} \right) + 2J$ \\
  \tabincell{c}{Log/Exp \\operations} & $MK{d_f}{I_t}\left( {{M^{{d_f} - 1}} + 1} \right) + 1$ &  0 \\
  \bottomrule
 \end{tabular}
\end{table*}

\begin{table*}[htbp]
\newcommand{\tabincell}[2]{\begin{tabular}{@{}#1@{}}#2\end{tabular}}  
 \caption{\label{TabNormedComlexity} Specific normalized complexity comparison}
 \centering
 \begin{tabular}{ccccc}
  \toprule
    & \tabincell{c}{Log-MPA $I_t=3$} & \tabincell{c}{Log-MPA $I_t=5$} & \tabincell{c}{Log-MPA $I_t=7$} & \tabincell{c}{DL-SCMA}\\
  \midrule
  Number of Mul. & 9456 & 9456 & 9456 & 14784\\
  Number of Add. & 13320 & 16920 & 20520 & 252\\
  Number of log/exp & 2449 & 4081 & 5713 & 0\\
  Normalized summation & 156860 & 193100 & 229340 & 148092\\
  \bottomrule
 \end{tabular}
\end{table*}

\begin{figure}
  \centering
    \includegraphics[width=0.8\textwidth]{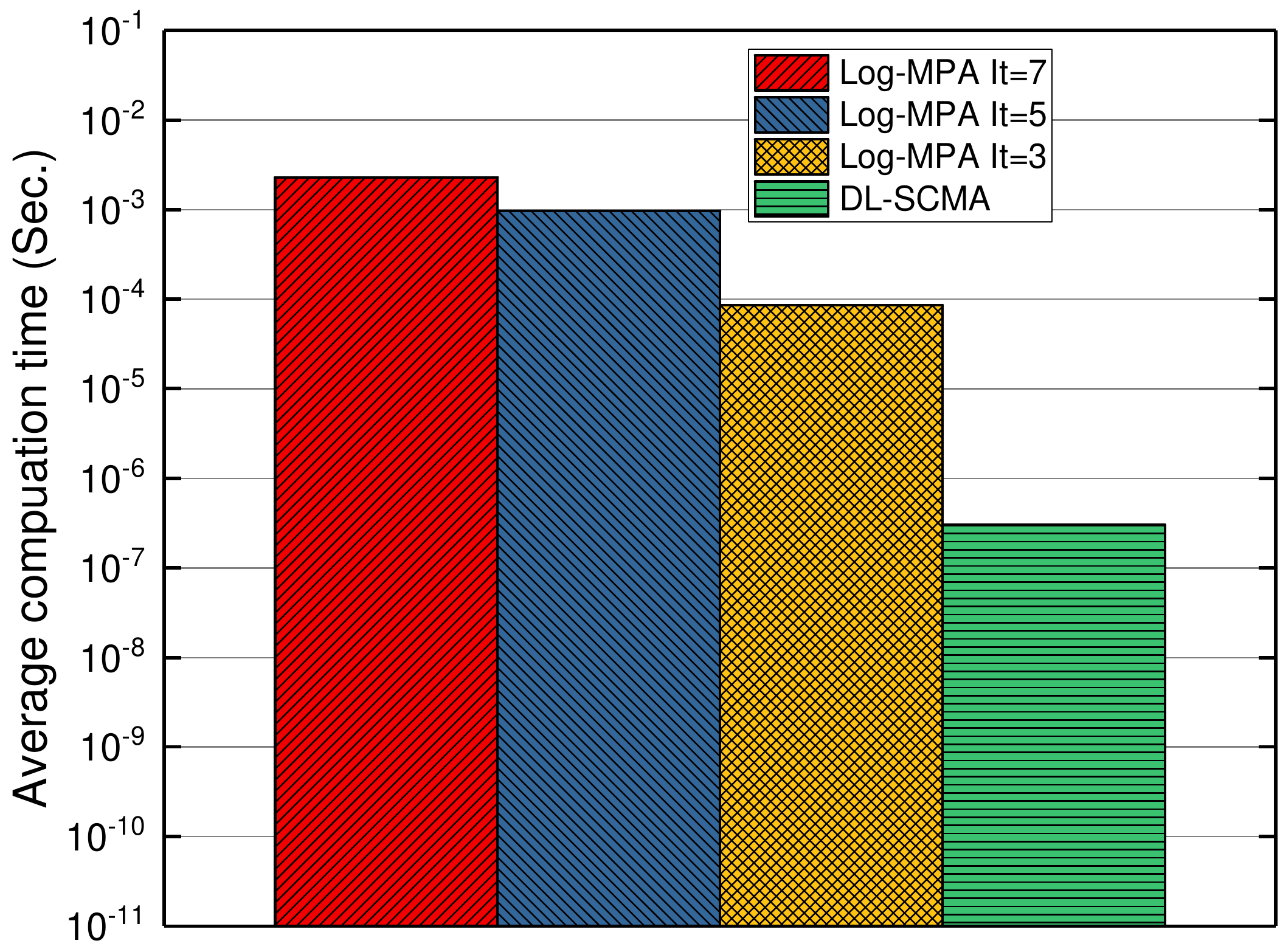}
  \caption{Average computation time comparison.}
  \label{CTBarGraph}
\end{figure}

Furthermore, we compare the actual run time on the host computer for the two approaches. Considering tensorflow may optimize the neural network and utilize GPU for acceleration, we have implemented the Log-MPA in tensorflow in the form of a static network model which has no trainable variable parameters. This can be done by a way similar to \cite{nachmani2018deep}, which expands the iterative message passing procedure in the belief propagation algorithm into stacked neural layers. For our case of implementing the pure Log-MPA for SCMA decoding in tensorflow, activation function is not used but \textit{tf.matmul}, \textit{tf.reduce\_logsumexp}, \textit{tf.log}, \textit{tf.exp} and some \textit{tf.unstack}, \textit{tf.concat}, \textit{tf.split} and \textit{tf.reshape} are took place. Fig. \ref{CTBarGraph} shows the average computational time for one received signal decoding consumed by different algorithms on the same host with the same software environment. It can be seen that DL-SCMA has about 2, 3 and 3 orders of magnitude improvements in saving computation time compare to Log-MPA at 3, 5 and 7 iterations, respectively.

\section{Discussions} \label{SecDis}
Users' codebooks for mapping transmitting bits into modulation constellation points are critical for SCMA performance. Traditionally, constellation structure analysis based codebook design methods are the mainstreams. They have succeeded in improving decoding accuracy for small-scale SCMA applications. However, when facing large-scale scenarios, for instance, when $M$, $N$, and $J$ are large enough, all of these methods might have difficulty in tackling the over-complexity optimal procedures. Through experiments given in this paper, it can be confirmed that autoencoders have the ability to learn to construct optimal SCMA codebooks. More importantly, an autoencoder is very easy to be established and expanded. A large number of large size codebooks learning can be achieved by simply enlarge the number of nodes of the input, output, and hidden layers. As a result, this could be a new way of thinking of SCMA codebook design methodology. We believe that other more DL techniques could also be valuable for SCMA codebook designing. Some extensive explorations should be aware of.

The DCMA scheme is the generalized version of SCMA. DCMA has the same overloading ratio as SCMA, as there are still $K$ resources mapping to $J$ users (layers). The difference between them lies in that DCMA maps every $m$ bits to $K$-dimensional codewords with $N' \approx K \gg N$ non-zero elements while SCMA has only $N \ll K$ non-zero elements. For a specific resource, overloading ratio $d_f$ is increased as more users' symbols are overlapping on it. However, for a specific user, information of its transmitting bits spread in more resources, so when decoding, more information can be utilized and more likely the transmitting bits can be decoded correctly. These two conditions are complementary, and it is the evidence of that the overall system overloading ratio remains unchanged. Although no extra information is brought in, DCMA could introduce different structure and organization in the received signal due to its dense mapping. By properly utilizing these new features DCMA provided, the decoding accuracy is expected to be improved. Compare to the MPA, DL-based methods are good at extracting new features from unknown information and utilizing them to accomplish the designed goal.

The reason why DCMA is not widely known and accepted is that it's hard to design a low complexity decoder, as the MPA has poor performance thus is not applicable. However, from the perspective of our proposed DNN decoder, all it's doing is learn to extract features suitable for constructing the decoding network from batches of training samples, so as to correctly decode new testing data. Thus, essentially, there is no difference for the DNN in decoding DCMA and SCMA. That's to say, by introducing DL based decoding method, it is no longer impossible to implement a feasible DCMA decoder. Taking advantage of DL technologies, we think that it should be given more concerns of the value of dense code based schemes, no matter DCMA, HDPC\cite{dimnik2009improved} or MDPC (high, moderate density parity check codes)\cite{chaulet2016worst}.

Finally, similar to the idea of this paper, perhaps other NOMA, e.g. power-domain NOMA, MUSA, and PDMA, can also utilize DL to improve their BER performances.

\section{Conclusions} \label{SecCon}
Motivated by the positive results of the emerging DL aided communication, we dedicated to exploring a novel DL based approach for SCMA systems, aiming to improve the BER performance. Firstly, we have proposed a DNN SCMA decoder which can be trained to decode all users' original transmitting bits without any prior knowledge of the users' codebooks and channel conditions. Further, an SCMA autoencoder is established which can automatically learn to construct optimal codebooks for all the users confronting AWGN and the corresponding decoder for recovering the original bits. Moreover, by changing the mapping vectors in the encoding part of AE-SCMA to dense vectors, a generalized SCMA named DCMA is established. Numerical simulations show the strong evidence that our proposed DL approach exhibits better BER and SER performance as well as sees lower computational complexity compared to the traditional Log-MPA.

With the aid of DL, designing a DNN for DCMA decoding become feasible, due to which DCMA scheme is considered reasonable and become an interesting topic. It is worth to pay significant effort for the in-depth research on this topic in the future. Meanwhile, software defined radio (SDR) experiments for evaluating the DNN SCMA decoder also remain to be implemented.

\section*{Acknowledgment}
This work was supported in part by the National Science Foundation of China (NSFC) under Grant 61433012 and Grant U1435215, in part by the Science Technology and Innovation Committee of Shenzhen Municipality under Grant GGFW2017073114031767, and in part by the Shenzhen Discipline Construction Project for Urban Computing and Data Intelligence.

\bibliographystyle{unsrt}

\begin{thebibliography}{10}

\bibitem{huang2019optimal}
Y.~Huang, J.~Wang, J.~Zhu, Optimal power allocation for downlink noma systems,
  in: Multiple Access Techniques for 5G Wireless Networks and Beyond, Springer,
  2019, pp. 195--227.

\bibitem{yuan2016multi}
Z.~Yuan, G.~Yu, W.~Li, Y.~Yuan, X.~Wang, J.~Xu, Multi-user shared access for
  {Internet of Things}, in: Vehicular Technology Conference (VTC Spring), 2016
  IEEE 83rd, IEEE, 2016, pp. 1--5.

\bibitem{chen2017pattern}
S.~Chen, B.~Ren, Q.~Gao, S.~Kang, S.~Sun, K.~Niu, Pattern division multiple
  access-a novel nonorthogonal multiple access for fifth-generation radio
  networks, IEEE Transactions on Vehicular Technology 66~(4) (2017) 3185--3196.

\bibitem{huang2014scalable}
J.~Huang, K.~Peng, C.~Pan, F.~Yang, H.~Jin, Scalable video broadcasting using
  bit division multiplexing, IEEE Transactions on Broadcasting 60~(4) (2014)
  701--706.

\bibitem{ping2006interleave}
L.~Ping, L.~Liu, K.~Wu, W.~K. Leung, Interleave division multiple-access, IEEE
  Transactions on Wireless Communications 5~(4) (2006) 938--947.

\bibitem{nikopour2013sparse}
H.~Nikopour, H.~Baligh, Sparse code multiple access, in: Personal Indoor and
  Mobile Radio Communications (PIMRC), 2013 IEEE 24th International Symposium
  on, IEEE, 2013, pp. 332--336.

\bibitem{hoshyar2008novel}
R.~Hoshyar, F.~P. Wathan, R.~Tafazolli, Novel low-density signature for
  synchronous {CDMA} systems over {AWGN} channel, IEEE Transactions on Signal
  Processing 56~(4) (2008) 1616--1626.

\bibitem{silver2016mastering}
D.~Silver, A.~Huang, C.~J. Maddison, A.~Guez, L.~Sifre, G.~Van Den~Driessche,
  J.~Schrittwieser, I.~Antonoglou, V.~Panneershelvam, M.~Lanctot, et~al.,
  Mastering the game of {Go} with deep neural networks and tree search, nature
  529~(7587) (2016) 484.

\bibitem{chen2018joint}
J.~Chen, Z.~Zhang, S.~Fu, J.~Hu, A joint update parallel mcmc-method-based
  sparse code multiple access decoder, IEEE Transactions on Vehicular
  Technology 67~(2) (2018) 1280--1291.

\bibitem{dai2017improved}
J.~Dai, K.~Niu, C.~Dong, J.~Lin, Improved message passing algorithms for sparse
  code multiple access, IEEE Transactions on Vehicular Technology 66~(11)
  (2017) 9986--9999.

\bibitem{tian2017low}
L.~Tian, M.~Zhao, J.~Zhong, P.~Xiao, L.~Wen, A low complexity detector for
  downlink {SCMA} systems, IET Communications 11~(16) (2017) 2433--2439.

\bibitem{Jia2018a}
M.~{Jia}, L.~{Wang}, Q.~{Guo}, X.~{Gu}, W.~{Xiang}, A low complexity detection
  algorithm for fixed up-link scma system in mission critical scenario, IEEE
  Internet of Things Journal 5~(5) (2018) 3289--3297.

\bibitem{wu2019low}
J.~Wu, S.~Wu, R.~Zuo, W.~Zhang, The low complexity multi-user detection
  algorithms for uplink scma system, in: International Conference on Wireless
  and Satellite Systems, Springer, 2019, pp. 562--575.

\bibitem{vameghestahbanati2017enabling}
M.~Vameghestahbanati, E.~Bedeer, I.~Marsland, R.~H. Gohary, H.~Yanikomeroglu,
  Enabling sphere decoding for {SCMA}, IEEE Communications Letters 21~(12)
  (2017) 2750--2753.

\bibitem{taherzadeh2014scma}
M.~Taherzadeh, H.~Nikopour, A.~Bayesteh, H.~Baligh, {SCMA} codebook design, in:
  Vehicular Technology Conference (VTC Fall), 2014 IEEE 80th, IEEE, 2014, pp.
  1--5.

\bibitem{zhou2017scma}
Y.~Zhou, Q.~Yu, W.~Meng, C.~Li, Scma codebook design based on constellation
  rotation, in: 2017 IEEE International Conference on Communications (ICC),
  IEEE, 2017, pp. 1--6.

\bibitem{bao2016spherical}
J.~Bao, Z.~Ma, M.~A. Mahamadu, Z.~Zhu, D.~Chen, Spherical codes for {SCMA}
  codebook, in: Vehicular Technology Conference (VTC Spring), 2016 IEEE 83rd,
  IEEE, 2016, pp. 1--5.

\bibitem{yu2018design}
L.~Yu, P.~Fan, D.~Cai, Z.~Ma, Design and analysis of scma codebook based on
  star-qam signaling constellations, IEEE Transactions on Vehicular Technology
  67~(11) (2018) 10543--10553.

\bibitem{peng2017joint}
J.~Peng, W.~Chen, B.~Bai, X.~Guo, C.~Sun, Joint optimization of constellation
  with mapping matrix for {SCMA} codebook design, IEEE Signal Processing
  Letters 24~(3) (2017) 264--268.

\bibitem{dong2018efficient2}
C.~Dong, X.~Cai, K.~Niu, J.~Lin, An efficient {SCMA} codebook design based on
  1-{D} searching algorithm, IEEE Communications Letters 22~(11) (2018)
  2234--2237.

\bibitem{sharma2018scma}
S.~Sharma, K.~Deka, V.~Bhatia, A.~Gupta, {SCMA} codebook based on optimization
  of mutual information and shaping gain, in: 2018 IEEE Globecom Workshops (GC
  Wkshps), IEEE, 2018, pp. 1--6.

\bibitem{dong2018efficient}
C.~Dong, G.~Gao, K.~Niu, J.~Lin, An efficient {SCMA} codebook optimization
  algorithm based on mutual information maximization, Wireless Communications
  and Mobile Computing 2018 (2018).

\bibitem{klimentyev2017scma}
V.~P. Klimentyev, A.~B. Sergienko, {SCMA} codebooks optimization based on
  genetic algorithm, in: European Wireless 2017; 23th European Wireless
  Conference; Proceedings of, VDE, 2017, pp. 1--6.

\bibitem{liu2018optimized}
S.~Liu, J.~Wang, J.~Bao, C.~Liu, Optimized {SCMA} codebook design by {QAM}
  constellation segmentation with maximized {MED}, IEEE Access 6 (2018)
  63232--63242.

\bibitem{xiao2018capacity}
K.~Xiao, B.~Xia, Z.~Chen, B.~Xiao, D.~Chen, S.~Ma, On capacity-based codebook
  design and advanced decoding for sparse code multiple access systems, IEEE
  Transactions on Wireless Communications (2018).

\bibitem{mheich2019design}
Z.~Mheich, L.~Wen, P.~Xiao, A.~Maaref, Design of {SCMA} codebooks based on
  golden angle modulation, IEEE Transactions on Vehicular Technology 68~(2)
  (2019) 1501--1509.

\bibitem{da2019multistage}
B.~F. da~Silva, D.~Silva, B.~F. Uch{\^o}a-Filho, D.~L. Ruyet, A multistage
  method for scma codebook design based on mds codes, arXiv preprint
  arXiv:1905.02533 (2019).

\bibitem{LIU201711}
W.~Liu, Z.~Wang, X.~Liu, N.~Zeng, Y.~Liu, F.~E. Alsaadi, A survey of deep
  neural network architectures and their applications, Neurocomputing 234
  (2017) 11 -- 26.

\bibitem{qin2019deep}
Z.~Qin, H.~Ye, G.~Y. Li, B.-H.~F. Juang, Deep learning in physical layer
  communications, IEEE Wireless Communications (2019).

\bibitem{o2017introduction}
T.~O'Shea, J.~Hoydis, An introduction to deep learning for the physical layer,
  IEEE Transactions on Cognitive Communications and Networking 3~(4) (2017)
  563--575.

\bibitem{felix2018ofdm}
A.~Felix, S.~Cammerer, S.~D{\"o}rner, J.~Hoydis, S.~t. Brink,
  {OFDM-Autoencoder} for end-to-end learning of communications systems, arXiv
  preprint arXiv:1803.05815 (2018).

\bibitem{huang2018deep}
H.~Huang, J.~Yang, H.~Huang, Y.~Song, G.~Gui, Deep learning for
  super-resolution channel estimation and doa estimation based massive mimo
  system, IEEE Transactions on Vehicular Technology 67~(9) (2018) 8549--8560.

\bibitem{huang2019deepmmware}
H.~Huang, Y.~Song, J.~Yang, G.~Gui, F.~Adachi, Deep-learning-based
  millimeter-wave massive mimo for hybrid precoding, IEEE Transactions on
  Vehicular Technology (2019).

\bibitem{karanov2018end}
B.~Karanov, M.~Chagnon, F.~Thouin, T.~A. Eriksson, H.~B{\"u}low, D.~Lavery,
  P.~Bayvel, L.~Schmalen, End-to-end deep learning of optical fiber
  communications, arXiv preprint arXiv:1804.04097 (2018).

\bibitem{lee2018deep}
H.~Lee, I.~Lee, S.~H. Lee, Deep learning based transceiver design for
  multi-colored {VLC} systems, Optics express 26~(5) (2018) 6222--6238.

\bibitem{nachmani2018deep}
E.~Nachmani, E.~Marciano, L.~Lugosch, W.~J. Gross, D.~Burshtein, Y.~Be’ery,
  Deep learning methods for improved decoding of linear codes, IEEE Journal of
  Selected Topics in Signal Processing 12~(1) (2018) 119--131.

\bibitem{liang2018iterative}
F.~Liang, C.~Shen, F.~Wu, An iterative {BP-CNN} architecture for channel
  decoding, IEEE Journal of Selected Topics in Signal Processing 12~(1) (2018)
  144--159.

\bibitem{west2017deep}
N.~E. West, T.~O'Shea, Deep architectures for modulation recognition, in:
  Dynamic Spectrum Access Networks (DySPAN), 2017 IEEE International Symposium
  on, IEEE, 2017, pp. 1--6.

\bibitem{kim2018novel}
M.~Kim, W.~Lee, D.-H. Cho, A novel {PAPR} reduction scheme for {OFDM} system
  based on deep learning, IEEE Communications Letters 22~(3) (2018) 510--513.

\bibitem{schmidt2017wireless}
M.~Schmidt, D.~Block, U.~Meier, Wireless interference identification with
  convolutional neural networks, arXiv preprint arXiv:1703.00737 (2017).

\bibitem{kim2018deep}
M.~Kim, N.-I. Kim, W.~Lee, D.-H. Cho, Deep learning aided {SCMA}, IEEE
  Communications Letters (2018).

\bibitem{liu201880}
Y.~Liu, Y.~Zhang, Low-dose ct restoration via stacked sparse denoising
  autoencoders, Neurocomputing 284 (2018) 80 -- 89.

\bibitem{CAO2018278}
W.~Cao, X.~Wang, Z.~Ming, J.~Gao, A review on neural networks with random
  weights, Neurocomputing 275 (2018) 278 -- 287.

\bibitem{glorot2010understanding}
X.~Glorot, Y.~Bengio, Understanding the difficulty of training deep feedforward
  neural networks, in: Proceedings of the thirteenth international conference
  on artificial intelligence and statistics, 2010, pp. 249--256.

\bibitem{wang2019batch}
J.~Wang, S.~Li, Z.~An, X.~Jiang, W.~Qian, S.~Ji, Batch-normalized deep neural
  networks for achieving fast intelligent fault diagnosis of machines,
  Neurocomputing 329 (2019) 53--65.

\bibitem{asiapresentation}
{Altera Innovate Asia website}, Presentation "1st 5g algorithm innovation
  competition-env1.0-scma",
  URL:http://www.innovateasia.com/5g/images/pdf/InnovateAsia\%20-\%201st\%205G\%20Algorithm\%20Competition\%20-\%20SCMA.pdf
  (2015).

\bibitem{brent2010modern}
R.~P. Brent, P.~Zimmermann, Modern computer arithmetic, Vol.~18, Cambridge
  University Press, 2010.

\bibitem{dimnik2009improved}
I.~Dimnik, Y.~Be'ery, Improved random redundant iterative {HDPC} decoding, IEEE
  Transactions on Communications 57~(7) (2009).

\bibitem{chaulet2016worst}
J.~Chaulet, N.~Sendrier, Worst case {QC-MDPC} decoder for {McEliece}
  cryptosystem, in: Information Theory (ISIT), 2016 IEEE International
  Symposium on, IEEE, 2016, pp. 1366--1370.


\end{thebibliography}

\end{document}